\documentclass[%
 aip,
 amsmath,amssymb,
 reprint,%
]{revtex4-1}

\usepackage{graphicx}% Include figure files
\graphicspath{ {./pictures/} }
\usepackage{dcolumn}% Align table columns on decimal point
\usepackage{bm}% bold math
\usepackage{tabularx}
\usepackage[utf8]{inputenc}
\usepackage[T1]{fontenc}
\usepackage{mathptmx}
\usepackage{etoolbox}
\usepackage{tikz}
\usepackage{amsmath}
\usepackage{multirow}
\usepackage{physics}
\usepackage[version=4]{mhchem}
\usepackage{adjustbox}
\makeatletter
\def\@email#1#2{%
 \endgroup
 \patchcmd{\titleblock@produce}
  {\frontmatter@RRAPformat}
  {\frontmatter@RRAPformat{\produce@RRAP{*#1\href{mailto:#2}{#2}}}\frontmatter@RRAPformat}
  {}{}
}%

\makeatother
\begin{document}

\preprint{AIP/123-QED}

\title[]{Toward Reliable Dipole Moments without Single Excitations: The Role of Orbital Rotations and Dynamical Correlation \newline}

\author{Rahul Chakraborty}
\author{Matheus Morato F. de Moraes}
\author{Katharina Boguslawski}
\author{Artur Nowak}
\affiliation{Institute of Physics, Faculty of Physics, Astronomy, and Informatics, Nicolaus Copernicus University in Toruń, Grudziądzka 5, 87-100 Torun, Poland }
\author{Julian Świerczyński}
\affiliation{Institute of Engineering and Technology, Faculty of Physics, Astronomy, and Informatics, Nicolaus Copernicus University in Toruń, Grudziądzka 5, 87-100 Toruń, Poland}

 \author{Paweł Tecmer}
\affiliation{Institute of Physics, Faculty of Physics, Astronomy, and Informatics, Nicolaus Copernicus University in Toruń, Grudziądzka 5, 87-100 Torun, Poland }
 
\email{matheusmorat@gmail.com, k.boguslawski@fizyka.umk.pl, ptecmer@fizyka.umk.pl, }

%\date{\today}% It is always \today, today,
             %  but any date may be explicitly specified

\begin{abstract}

Abstract: The dipole moment is a crucial molecular property linked to a molecular system's bond polarity and overall electronic structure.
To that end, the electronic dipole moment, which results from the electron density of a system, is often used to assess the accuracy and reliability of new electronic structure methods. 
This work analyses electronic dipole moments computed with the pair coupled cluster doubles (pCCD) ansatz and its linearized coupled cluster (pCCD-LCC) corrections using the canonical Hartree--Fock and pCCD-optimized (localized) orbital bases.
The accuracy of pCCD-based dipole moments is assessed against experimental and CCSD(T) reference values using relaxed and unrelaxed density matrices and different basis set sizes. 
Our test set comprises molecules of various bonding patterns and electronic structures, exposing pCCD-based methods to a wide range of electron correlation effects.
Additionally, we investigate the performance of pCCD-in-DFT dipole moments of some model complexes. 
Finally, our work indicates the importance of orbital relaxation in the pCCD model and shows the limitations of the linearized couple cluster corrections in predicting electronic dipole moments of multiple-bonded systems.
Most importantly, pCCD with a linearized CCD correction can reproduce the dipole moment surfaces in singly-bonded molecules, which are comparable to the multi-reference ones.
\end{abstract}
\maketitle 

%%%%%%%%%%%%%%%%%%%%%%%%%%%%%%%%%%%%%%%%%%%%%%%%%%%%%%%%%%%%%%%%%%%%%
%% Start the main part of the manuscript here.
%%%%%%%%%%%%%%%%%%%%%%%%%%%%%%%%%%%%%%%%%%%%%%%%%%%%%%%%%%%%%%%%%%%%%
%%%%%%%%%%%%%%%%%%%%%%%%%%%%%%%%%%%%%%%%%%%%%%%%%%%%%%%%%%%%%%%%%%%%%
%%% Introduction
%%%%%%%%%%%%%%%%%%%%%%%%%%%%%%%%%%%%%%%%%%%%%%%%%%%%%%%%%%%%%%%%%%%%% 
\section{\label{sec:level1}Introduction}

The electric dipole moment is the major component of electrostatic interactions, which plays a significant role in many areas of chemistry, physics, and biology.~\cite{honig1995classical}
The electronic component of the molecular dipole moment contains many finer details about the electronic structure and bonding patterns in molecules~\cite{frenking2007electronic} and contributes to interpreting spectroscopic data.~\cite{neugebauer2002fullerene, fedorov2017ab} Dipole moment surfaces, on the other hand, provide information about the change in bond polarity,~\cite{buldakov2009_dipole-sign} intensities of the rovibrational transitions~\cite{lodi2008new} etc.  
The reliable determination of this fundamental property is, thus, of preliminary importance for both experimental and theoretical domains. 
To that end, the quantum chemical modeling of electronic dipole moment provides a common testing ground for approximate wave function models.~\cite{cohen1965electric, yoshimine1967ground, green1971electric, boyd1972electron, diercksen1983perturbation} 
They can be compared with experimental results that are readily available for many small molecules. 
For example, the dipole moment was benchmarked against quantum chemical methods like Hartree--Fock theory, second-order M{\o}ller--Plesset (MP2) perturbation theory, coupled-cluster (CC) methods, multi-reference methods, and density functional theory (DFT) approximations.~\cite{hickey2014benchmarking, liu2023ccsdtperformance} 
Specifically, coupled cluster-based ans\"atze have been extensively tested for dipole moment properties~\cite{noga1988expectation, korona2004effect} and remain an active research field.~\cite{bozkaya2020state, traore2022basis} 
Maroulis and coworkers~\cite{maroulis1998systematic, maroulis1988multipole, maroulis1988multipoleerratum, maroulis1985electric_BH, maroulis1992electric_CS2, maroulis1991hyperpolarizability, maroulis2003electric_HF} performed numerous coupled cluster calculations, including the quantum chemistry gold standard---coupled cluster singles and doubles with perturbative triples (CCSD(T)), on electronic properties of different system types ranging from small di- and triatomic to organic molecules.~\cite{maroulis2001dipole_adamantane}
Studies by Mazziotti and coworkers~\cite{gidofalvi2006computation, gidofalvi2007molecular, mazziotti2010parametrization} have shown alternate routes for evaluation of electric properties using variational reduced density matrices.
The elimination of the need for any reference wave function in this approach has great promise for determining electric property in systems with multi-reference characters.

Although the electric dipole moment can be easily determined through density matrices, its sensitivity toward the accuracy of the electron density poses a real challenge to various quantum chemical methods.~\cite{korona-S-operator-CC-jcp-2006, fecteau2020reduced, johnson2021transition}
First, orbital relaxation has been shown to have a profound role in this regard.~\cite{salter1987property, korona2004effect, kats2014communication}
Second, some molecules require the inclusion of triple (or higher) excitations in the wave function expansion to obtain reliable dipole moments.~\cite{klopper1997multiple, liu2023performance} 
The above aspects are the source of the well-known struggle approximate quantum chemistry methods face in an accurate description of the dipole moment of the CO molecule. \cite{kello1988CO, werner1981mcscf, scuseria1991dipole, barnes1993bond, schautz1999montecarlo_CO, buldakov2009_dipole-sign, meshkov2022semi}

There are new families of geminal-based methods~\cite{hapka-jensen,  johnson_2013, johnson2017strategies, fency, johnson2020richardson, johnson2021transition, johnson2022bivariational} that are yet to be thoroughly tested for dipole moment properties. 
Some of the most promising ones are those based on the pCCD ansatz.~\cite{limacher-ap1rog-jctc-2013,oo-ap1rog,tamar-pccd, pawel-pccp-geminal-review-2022}
They have seen recent successes in treating strongly correlated systems with mean-field-like scaling. 
pCCD has the feature of using its optimized orbital basis without defining active spaces.~\cite{oo-ap1rog, ps2-ap1rog, piotrus_mol-phys, ap1rog-non-variational-orbital-optimizarion-jctc} 
The size-extensive and size-consistent nature of orbital-optimized pCCD has motivated a wide range of studies for covalent molecules,~\cite{pawel_jpca_2014, ps2-ap1rog, piotrus_mol-phys, ap1rog-non-variational-orbital-optimizarion-jctc, pawel-pccp-2015, garza2015actinide, pccd-prb-2016, pybest-paper, ola-tcc, ola-qit-actinides-pccp-2022} non-covalent systems,~\cite{filip-jctc-2019, garza-pccp-wdv} and excited states,~\cite{eom-pccd, eom-pccd-erratum, state-specific-oopccd, ip-pccd} including organic systems.~\cite{pccd-delaram-rsc-adv-2023, pccd-perspective-jpcl-2023} 
Perturbation theory-based, conventional and linearized coupled cluster (LCC) corrections have also been successfully added to the pCCD wave function to improve the description of dynamic correlation.~\cite{piotrus_pt2, frozen-pccd, eom-pccd-lccsd, pccd-ptx, piotrus-orbital-energies, pccd-ee-f0-actinides} 

To the best of our knowledge, little is known about the performance of the pCCD family of methods for ground-state electronic properties like dipole moments.
However, there have been studies for such properties with the antisymmetric product of strongly orthogonal geminals (APSG)~\cite{apsg-cndo-dipole-mp-1975, apsg-vs-mcscf-dipole-jcp-1980} and other pair-coupled approximate wave function methods.
The natural orbital functional theory formulated by Piris and coworkers (PNOFi, i=1,6) is noteworthy in this respect.~\cite{piris2013natural, piris2014interacting} 
Specifically, the PNOF5 is similar to the APSG approach~\cite{pernal2013equivalence} and, thus, indirectly related to pCCD.~\cite{pawel-pccp-geminal-review-2022}
The coupled electron pair approximation(0) (CEPA(0)) and its orbital optimized variant~\cite{bozkaya2013orbital} have similarities with the LCC approach. 
CEPA-based methods have been tested for dipole moments of various molecules.\cite{meyer1975pno, werner1976pno, werner1980theoretical}

This work aims to assess the performance of pCCD-type methods in quantifying the electric dipole moments of 
diatomics of the main group elements, the first transition metal series, and some larger complexes. 
The selected diatomic systems represent various bonding patterns (metal--nonmetal, nonmetal--nonmetal, metalloid--nonmetal, metal--metal van der Waals interaction).
However, the pCCD framework restricts us to molecules with singlet ground states. 
Our work focuses on the effects of orbital optimization within pCCD and the inclusion of dynamic correlation. 
We use linearized coupled-cluster methods for the latter on top of the Hartree--Fock and pCCD wave function: doubles (pCCD-LCCD) and singles and doubles (pCCD-LCCSD) models.~\cite{ap1rog-lcc, post-pccd-entanglement} 
Furthermore, we probe the sensitivity of pCCD-based methods for dipole moments regarding different basis set sizes. 
We compare our electronic dipole moment values with the CCSD and CCSD(T) methods using relaxed and unrelaxed density matrices and experimental values.
Finally, we extend our studies to pCCD-based static embedding calculations, where we obtain the embedding potential through the DFT approach (pCCD-in-DFT).~\cite{gomes_rev_2012, pccd-static-embedding} 
Precisely, we assess the performance of the pCCD-in-DFT embedding model for the electronic dipole moments of weakly hydrogen-bonded binary complexes such as \ce{CO-HF}, \ce{CO-HCl}, \ce{N2-HF}, \ce{N2-HCl}, and the \ce{H2O \cdots Rg} [Rg = He, Ne, Ar, Kr] van der Waals complexes.
The electronic structures of these complexes have been studied with various quantum chemical methods and thus represent a good reference point.~\cite{curtiss1985investigation, bacskay1990quantum, woon1996_n2-hf, haskopoulos2010interaction}
Additionally, the weak interactions present in these molecules provide a good testing ground for the static embedding approach.
%We benchmark the results from the static embedding approach against supramolecular pCCD values.
In summary, this work reports the performance of some unique pCCD-based models (with and without orbital optimization) with and without dynamic energy corrections for dipole moment calculations. 

%%%%%%%%%%%%%%%%%%%%%%%%%%%%%%%%%%%%%%%%%%%%%%%%%%%%%%%%%%%%%%%%%%%%%
%%%%%%%%%%%%  Theory pCCD methods      %%%%%%%%%%%%%%%%%%%%%%%%%%%%%%
%%%%%%%%%%%%%%%%%%%%%%%%%%%%%%%%%%%%%%%%%%%%%%%%%%%%%%%%%%%%%%%%%%%%% 
\section{Theory}
%%%%%%%%%%%%%%%%%%%%%%%%%%%%%%%%%%%%%%%%%%%%%%%%%%%%%%%%%%%%%%%%%%%%%
\subsection{pCCD and Related Methods}
%%%%%%%%%%%%%%%%%%%%%%%%%%%%%%%%%%%%%%%%%%%%%%%%%%%%%%%%%%%%%%%%%%%%%
Limiting the cluster operator to pair-excitations in the coupled cluster ansatz produces the pCCD ansatz, 
%%%%%%%%%%%%%%%%%%%%%%%%%%%%%%%%%%%%%%%%%%%%%%%%%%%%%%%%%%%%%%%%%%
%%%%%%%%%%%%%%%%%%  Eq. 1 (pCCD)    %%%%%%%%%%%%%%%%%%%%%%%%%%%%%%
%%%%%%%%%%%%%%%%%%%%%%%%%%%%%%%%%%%%%%%%%%%%%%%%%%%%%%%%%%%%%%%%%%
\begin{align}
     \label{eqn:pccd}
     \ket{\Psi_{\textrm{pCCD}}} = 
        \textrm{exp}\left(\sum_{i}^{\rm occ}\sum_{a}^{\rm virt}t_{i}^{a}\hat{a}_{{a}}^{\dagger}\hat{a}_{\bar{a}}^{\dagger}\hat{a}_{{\bar{i}}}\hat{a}_{i}\right)\ket {0} 
        = \textrm{e}^{\hat{T}_{\rm p}}\ket {0},
\end{align}
%%%%%%%%%%%%%%%%%%%%%%%%%%%%%%%%%%%%%%%%%%%%%%%%%%%%%%%%%%%%%%%%%%
%%%%%%%%%%%%%%%%%%  End of Eq. 1   %%%%%%%%%%%%%%%%%%%%%%%%%%%%%%%
%%%%%%%%%%%%%%%%%%%%%%%%%%%%%%%%%%%%%%%%%%%%%%%%%%%%%%%%%%%%%%%%%%
where $\hat {a}_{p}^{\dagger}$ and $\hat {a}_{p}$ ($\hat {a}_{\bar{p}}^{\dagger}$ and $\hat {a}_{\bar{p}}$) are the creation and annihilation operators for $\alpha$-spin (and $\beta$-spin) electrons. 
$\hat T_{\rm p}$ is the pair-excitation cluster operator and $\ket{0}$ is a reference independent particle model, usually the Hartree--Fock wave function.
The pCCD model misses a significant fraction of the dynamic electron correlation effects. 
In this work, we use a posteriori linearized coupled cluster~\cite{ap1rog-lcc} (LCC) corrections on top of the pCCD wave function to compensate for that. 
In the LCC correction, the exponential coupled cluster ansatz with a pCCD reference wave function is used as
%%%%%%%%%%%%%%%%%%%%%%%%%%%%%%%%%%%%%%%%%%%%%%%%%%%%%%%%%%%%%%%%%%
%%%%%%%%%%%%%%%%%%  Eq. 2 (LCC)     %%%%%%%%%%%%%%%%%%%%%%%%%%%%%%
%%%%%%%%%%%%%%%%%%%%%%%%%%%%%%%%%%%%%%%%%%%%%%%%%%%%%%%%%%%%%%%%%%
\begin{equation}
\label{eqn:lcc}
\ket{\Psi}  =  \exp(\hat{T}') \ket{\Psi_{\rm pCCD}},
\end{equation}
%%%%%%%%%%%%%%%%%%%%%%%%%%%%%%%%%%%%%%%%%%%%%%%%%%%%%%%%%%%%%%%%%%
%%%%%%%%%%%%%%%%%%  End of Eq. 2   %%%%%%%%%%%%%%%%%%%%%%%%%%%%%%%
%%%%%%%%%%%%%%%%%%%%%%%%%%%%%%%%%%%%%%%%%%%%%%%%%%%%%%%%%%%%%%%%%%
where $\hat{T}'=\sum_{\nu}t_{\nu}\hat{\tau}_{\nu}$ is a cluster operator containing excitation operators $\hat{\tau}_{\nu}$ of various levels.
The "$\prime$" in the cluster operator indicates that the pair excitations present in pCCD are excluded.
The corresponding energy equation is
%%%%%%%%%%%%%%%%%%%%%%%%%%%%%%%%%%%%%%%%%%%%%%%%%%%%%%%%%%%%%%%%%%
%%%%%%%%%%%%%%%%%%  Eq. 3 (LCC Ham)     %%%%%%%%%%%%%%%%%%%%%%%%%%
%%%%%%%%%%%%%%%%%%%%%%%%%%%%%%%%%%%%%%%%%%%%%%%%%%%%%%%%%%%%%%%%%%
\begin{equation}
    \hat{H}\exp(\hat{T}') \ket{\Psi_{\rm pCCD}} = E\exp(\hat{T}') \ket{\Psi_{\rm pCCD}},
\end{equation}
%%%%%%%%%%%%%%%%%%%%%%%%%%%%%%%%%%%%%%%%%%%%%%%%%%%%%%%%%%%%%%%%%%
%%%%%%%%%%%%%%%%%%  End of Eq. 3   %%%%%%%%%%%%%%%%%%%%%%%%%%%%%%%
%%%%%%%%%%%%%%%%%%%%%%%%%%%%%%%%%%%%%%%%%%%%%%%%%%%%%%%%%%%%%%%%%%
where $\hat{H}$ is the electronic Hamiltonian of the system.
In the LCC framework, the associated Baker--Campbell--Hausdorff expansion is restricted to the second term, i.e.,
%%%%%%%%%%%%%%%%%%%%%%%%%%%%%%%%%%%%%%%%%%%%%%%%%%%%%%%%%%%%%%%%%%
%%%%%%%%%%%%%%%%%%  Eq. 4 (LCC BCH)     %%%%%%%%%%%%%%%%%%%%%%%%%%
%%%%%%%%%%%%%%%%%%%%%%%%%%%%%%%%%%%%%%%%%%%%%%%%%%%%%%%%%%%%%%%%%%
\begin{equation}
    (\hat{H}+[\hat{H},\hat{T}']) \ket{\Psi_{\rm pCCD}} = E\ket{\Psi_{\rm pCCD}}.
\end{equation}
%%%%%%%%%%%%%%%%%%%%%%%%%%%%%%%%%%%%%%%%%%%%%%%%%%%%%%%%%%%%%%%%%%
%%%%%%%%%%%%%%%%%%  End of Eq. 4   %%%%%%%%%%%%%%%%%%%%%%%%%%%%%%%
%%%%%%%%%%%%%%%%%%%%%%%%%%%%%%%%%%%%%%%%%%%%%%%%%%%%%%%%%%%%%%%%%%
When we include both single and double excitations, $\hat{T}'$ reads,
%%%%%%%%%%%%%%%%%%%%%%%%%%%%%%%%%%%%%%%%%%%%%%%%%%%%%%%%%%%%%%%%%%
%%%%%%%%%%%%%%%%%%  Eq. 5 (T')          %%%%%%%%%%%%%%%%%%%%%%%%%%
%%%%%%%%%%%%%%%%%%%%%%%%%%%%%%%%%%%%%%%%%%%%%%%%%%%%%%%%%%%%%%%%%%
\begin{equation}
\label{eqn:lccsd}
{\hat{T}'= \hat T_1 + \hat T_2'= \sum_{i}^{\rm occ} \sum_{a}^{\rm virt} t_{i}^a \hat{E}_{ai} + \frac{1}{2}\sum_{i, j}^{\rm occ}~ \sum_{a, b}^{\rm virt}{'}~t_{ij}^{ab}~\hat{E}_{ai}\hat{E}_{bj} }, 
\end{equation}
%%%%%%%%%%%%%%%%%%%%%%%%%%%%%%%%%%%%%%%%%%%%%%%%%%%%%%%%%%%%%%%%%%
%%%%%%%%%%%%%%%%%%  End of Eq. 5   %%%%%%%%%%%%%%%%%%%%%%%%%%%%%%%
%%%%%%%%%%%%%%%%%%%%%%%%%%%%%%%%%%%%%%%%%%%%%%%%%%%%%%%%%%%%%%%%%%
where
%%%%%%%%%%%%%%%%%%%%%%%%%%%%%%%%%%%%%%%%%%%%%%%%%%%%%%%%%%%%%%%%%%
%%%%%%%%%%%%%%%%%%  Eq. 6 (E_ai)        %%%%%%%%%%%%%%%%%%%%%%%%%%
%%%%%%%%%%%%%%%%%%%%%%%%%%%%%%%%%%%%%%%%%%%%%%%%%%%%%%%%%%%%%%%%%%
\begin{equation}
\label{eqn:q_ai}
{\hat{E}_{ai} = \hat{a}_{a}^{\dagger}\hat{a}_{i} + \hat{a}_{\bar{a}}^{\dagger}\hat{a}_{\bar{i}}}
\end{equation}   
%%%%%%%%%%%%%%%%%%%%%%%%%%%%%%%%%%%%%%%%%%%%%%%%%%%%%%%%%%%%%%%%%%
%%%%%%%%%%%%%%%%%%  End of Eq. 6   %%%%%%%%%%%%%%%%%%%%%%%%%%%%%%%
%%%%%%%%%%%%%%%%%%%%%%%%%%%%%%%%%%%%%%%%%%%%%%%%%%%%%%%%%%%%%%%%%%
is the singlet excitation operator.
Note that the "$\prime$" in the second sum of the above equation excludes the cases where $i=j$ and simultaneously $a=b$, while terms where $i=j \wedge a\neq b$ or $i \ne j \wedge a = b$ are still included.
Elimination of $\hat{T_{1}}$ amplitudes from $\hat{T}'$ in eq.~\ref{eqn:lccsd} leads to the pCCD-LCCD model. 
Both pCCD-LCC variants have been successfully used for various molecules, providing a moderate balance between dynamic and non-dynamic electron correlation effects.~\cite{ap1rog-lcc, filip-jctc-2019, pawel-yb2, pccd-ptx, ola-tcc}
%%%%%%%%%%%%%%%%%%%%%%%%%%%%%%%%%%%%%%%%%%%%%%%%%%%%%%%%%%%%%%%%%%
%%%%%%%%%%%%  Theory: DMs              %%%%%%%%%%%%%%%%%%%%%%%%%%%
%%%%%%%%%%%%%%%%%%%%%%%%%%%%%%%%%%%%%%%%%%%%%%%%%%%%%%%%%%%%%%%%%% 
\subsection{Density Matrices from pCCD and Related Methods}
%%%%%%%%%%%%%%%%%%%%%%%%%%%%%%%%%%%%%%%%%%%%%%%%%%%%%%%%%%%%%%%%%% 
Elements of the 1-electron reduced density matrix (1-RDM) obtained from any wave function $\Psi$ can be expressed as
%%%%%%%%%%%%%%%%%%%%%%%%%%%%%%%%%%%%%%%%%%%%%%%%%%%%%%%%%%%%%%%%%%
%%%%%%%%%%%%%%%%%%  Eq. 7 (1-DM)        %%%%%%%%%%%%%%%%%%%%%%%%%%
%%%%%%%%%%%%%%%%%%%%%%%%%%%%%%%%%%%%%%%%%%%%%%%%%%%%%%%%%%%%%%%%%%
\begin{equation}
\label{eq:1-RDM}
    \gamma_{q}^{p}= \bra{\Psi}a_{p}^{\dagger}a_{q}\ket{\Psi}.
\end{equation}
%%%%%%%%%%%%%%%%%%%%%%%%%%%%%%%%%%%%%%%%%%%%%%%%%%%%%%%%%%%%%%%%%%
%%%%%%%%%%%%%%%%%%  End of Eq. 7   %%%%%%%%%%%%%%%%%%%%%%%%%%%%%%%
%%%%%%%%%%%%%%%%%%%%%%%%%%%%%%%%%%%%%%%%%%%%%%%%%%%%%%%%%%%%%%%%%%
For truncated CC models, the 1-electron molecular response properties are calculated using the derivative approach as a response to a small external perturbation related to the property in question (such as dipole moments). 
In this approach, the response density matrices are often used.~\cite{salter1989analytic, Kraka1991responseDM, bokhan2016electric, oo-lccd-lambda-equations-pccp-2016}
Accordingly, elements of the pCCD response 1-RDM are defined as 
%%%%%%%%%%%%%%%%%%%%%%%%%%%%%%%%%%%%%%%%%%%%%%%%%%%%%%%%%%%%%%%%%%
%%%%%%%%%%%%%%%%%%  Eq. 8 (1-RDM)       %%%%%%%%%%%%%%%%%%%%%%%%%%
%%%%%%%%%%%%%%%%%%%%%%%%%%%%%%%%%%%%%%%%%%%%%%%%%%%%%%%%%%%%%%%%%%
\begin{equation}
\label{eq:response-1-RDM}
    ^{\rm pCCD}\gamma_{q}^{p}= \bra {0}(1+\Lambda_{\rm pCCD})\textrm{e}^{-\hat{T}_{\rm pCCD}}{{\hat{a}_{p}^{\dagger}\hat{a}_{q}}}\textrm{e}^{\hat{T}_{\rm pCCD}}\ket {0},
\end{equation}
%%%%%%%%%%%%%%%%%%%%%%%%%%%%%%%%%%%%%%%%%%%%%%%%%%%%%%%%%%%%%%%%%%
%%%%%%%%%%%%%%%%%%  End of Eq. 8   %%%%%%%%%%%%%%%%%%%%%%%%%%%%%%%
%%%%%%%%%%%%%%%%%%%%%%%%%%%%%%%%%%%%%%%%%%%%%%%%%%%%%%%%%%%%%%%%%%
where $\Lambda_{\rm pCCD}=\sum_{ia}\lambda_{a}^{i}\hat{a}_{{i}}^{\dagger}\hat{a}_{\bar{i}}^{\dagger}\hat{a}_{{\bar{a}}}\hat{a}_{a}$ is the electron-pair de-excitation operator. 

On the other hand, the response 1-RDM from the pCCD-LCC wave functions can be constructed using the reference response 1-RDM of pCCD from eq.~\ref{eq:response-1-RDM} and the correlation contribution of the LCC correction on top of the pCCD wave function calculated using the so-called $\Lambda$-equations,~\cite{post-pccd-entanglement}
%%%%%%%%%%%%%%%%%%%%%%%%%%%%%%%%%%%%%%%%%%%%%%%%%%%%%%%%%%%%%%%%%%
%%%%%%%%%%%%%%%%%%  Eq. 9  %%%%%%%%%%%%%%%%%%%%%%%%%%%%%%%%%%%%%%%
%%%%%%%%%%%%%%%%%%%%%%%%%%%%%%%%%%%%%%%%%%%%%%%%%%%%%%%%%%%%%%%%%%
\begin{align}\label{eq:responsedms-corr}
    ^{\rm LCC}\gamma^{p}_{q}=\bra{0}(1+\Lambda^\prime_{\rm LCC}) & \{ e^{-\hat{T}^\prime -\hat{T}_\textrm{pCCD}} \{\hat{a}^\dagger_p a_q \}  e^{\hat{T}_\textrm{pCCD}+\hat{T}^\prime} \}_{L^\prime} \ket{0},
\end{align}
%%%%%%%%%%%%%%%%%%%%%%%%%%%%%%%%%%%%%%%%%%%%%%%%%%%%%%%%%%%%%%%%%%
%%%%%%%%%%%%%%%%%%  End of Eq. 9   %%%%%%%%%%%%%%%%%%%%%%%%%%%%%%%
%%%%%%%%%%%%%%%%%%%%%%%%%%%%%%%%%%%%%%%%%%%%%%%%%%%%%%%%%%%%%%%%%%
where $\Lambda^\prime_{\rm LCC}=\Lambda_1+\Lambda_2^\prime$ or $\Lambda^\prime_{\rm LCC}=\Lambda_2^\prime$, respectively, and
%%%%%%%%%%%%%%%%%%%%%%%%%%%%%%%%%%%%%%%%%%%%%%%%%%%%%%%%%%%%%%%%%%
%%%%%%%%%%%%%%%%%%  Eq. 10 %%%%%%%%%%%%%%%%%%%%%%%%%%%%%%%%%%%%%%%
%%%%%%%%%%%%%%%%%%%%%%%%%%%%%%%%%%%%%%%%%%%%%%%%%%%%%%%%%%%%%%%%%%
\begin{equation}\label{eq:de-excitation}
\Lambda_n^\prime = {\frac{1}{(n!)^2}}\sum_{ij\ldots}\sum_{ab\ldots}{}^\prime \lambda^{ij\ldots}_{ab\ldots}{i^\dagger a j^\dagger b\ldots}
\end{equation}
%%%%%%%%%%%%%%%%%%%%%%%%%%%%%%%%%%%%%%%%%%%%%%%%%%%%%%%%%%%%%%%%%%
%%%%%%%%%%%%%%%%%%  End of Eq. 10   %%%%%%%%%%%%%%%%%%%%%%%%%%%%%%
%%%%%%%%%%%%%%%%%%%%%%%%%%%%%%%%%%%%%%%%%%%%%%%%%%%%%%%%%%%%%%%%%%
is the de-excitation operator, where all electron-pair de-excitation are to be excluded as they do not enter the {LCC} equations (again, indicated by the "$\prime$").
For the {LCC} response density matrices, only terms that are at most linear in $\hat{T}_1$ and $\hat{T}_2^\prime$ are to be considered.
This is indicated by $\{\ldots \}_{L^\prime}$ in the above equation.
The final 1-RDM from oo-pCCD-LCC(S)D approaches is the sum of the relaxed oo-pCCD and unrelaxed LCC(S)D contributions,
%%%%%%%%%%%%%%%%%%%%%%%%%%%%%%%%%%%%%%%%%%%%%%%%%%%%%%%%%%%%%%%%%%
%%%%%%%%%%%%%%%%%%  Eq. 11 %%%%%%%%%%%%%%%%%%%%%%%%%%%%%%%%%%%%%%%
%%%%%%%%%%%%%%%%%%%%%%%%%%%%%%%%%%%%%%%%%%%%%%%%%%%%%%%%%%%%%%%%%%
\begin{equation}\label{eq:responsedms}
\gamma^{p}_{q}= {}^{\rm pCCD}\gamma_{q}^{p} + {}^{\rm LCC}\gamma_{q}^{p}.
\end{equation}
%%%%%%%%%%%%%%%%%%%%%%%%%%%%%%%%%%%%%%%%%%%%%%%%%%%%%%%%%%%%%%%%%%
%%%%%%%%%%%%%%%%%%  End of Eq. 11  %%%%%%%%%%%%%%%%%%%%%%%%%%%%%%%
%%%%%%%%%%%%%%%%%%%%%%%%%%%%%%%%%%%%%%%%%%%%%%%%%%%%%%%%%%%%%%%%%%
As evident from eq.~\ref{eq:response-1-RDM}, ${}^{\rm pCCD}\gamma_{q}^{p}$ contains both the contribution from the reference determinant and the pCCD-correlation part, ${}^{\rm pCCD}\gamma_{q}^{p} = {}^{ref}\gamma_{q}^{p} +{}^{{\rm corr}({\rm pCCD})}\gamma_{q}^{p}$, while ${}^{\rm LCC}\gamma_{q}^{p}$ accounts for the LCC correlation part only.
It is to be noted that orbital relaxation due to the LCC correction is not considered in this work. 

%%%%%%%%%%%%%%%%%%%%%%%%%%%%%%%%%%%%%%%%%%%%%%%%%%%%%%%%%%%%%%%%%%
%%%%%%%%%%%%  Theory: Dipole Moment     %%%%%%%%%%%%%%%%%%%%%%%%%%
%%%%%%%%%%%%%%%%%%%%%%%%%%%%%%%%%%%%%%%%%%%%%%%%%%%%%%%%%%%%%%%%%% 
\subsection{Dipole moment calculation}
%%%%%%%%%%%%%%%%%%%%%%%%%%%%%%%%%%%%%%%%%%%%%%%%%%%%%%%%%%%%%%%%%% 
The total dipole moment of a molecule is defined as 
%%%%%%%%%%%%%%%%%%%%%%%%%%%%%%%%%%%%%%%%%%%%%%%%%%%%%%%%%%%%%%%%%%
%%%%%%%%%%%%%%%%%%  Eq. 12 (dipolr_u)   %%%%%%%%%%%%%%%%%%%%%%%%%%
%%%%%%%%%%%%%%%%%%%%%%%%%%%%%%%%%%%%%%%%%%%%%%%%%%%%%%%%%%%%%%%%%%
\begin{align}
\label{eq:dipole_def_total}
    \mu_{\alpha} = \sum_{i=1}^{N_{\rm nuc}}Z_{i}R_{i\alpha} -\int{\rho(r)r_{\alpha}dr},
\end{align}
%%%%%%%%%%%%%%%%%%%%%%%%%%%%%%%%%%%%%%%%%%%%%%%%%%%%%%%%%%%%%%%%%%
%%%%%%%%%%%%%%%%%%  End of Eq. 12   %%%%%%%%%%%%%%%%%%%%%%%%%%%%%%%
%%%%%%%%%%%%%%%%%%%%%%%%%%%%%%%%%%%%%%%%%%%%%%%%%%%%%%%%%%%%%%%%%%
where the first term accounts for nuclear and the second for electronic contributions. 
In eq.~\ref{eq:dipole_def_total}, $\alpha$ denotes the axial direction (x, y or z), $Z_{i}$ charge of the i-th nucleus, $N_{\rm nuc}$ the number of nuclei in the molecular structure, and $R$ and $r$ correspond to the nuclear and electronic coordinates, respectively. 

%In quantum chemical calculations, the electron density of the molecular structure $\rho(r)$ in eq.~\ref{eq:dipole_def_total} can be represented 
%by the density matrix, 
%%%%%%%%%%%%%%%%%%%%%%%%%%%%%%%%%%%%%%%%%%%%%%%%%%%%%%%%%%%%%%%%%%
%%%%%%%%%%%%%%%%%%  Eq. 13 (rho_MO)     %%%%%%%%%%%%%%%%%%%%%%%%%%
%%%%%%%%%%%%%%%%%%%%%%%%%%%%%%%%%%%%%%%%%%%%%%%%%%%%%%%%%%%%%%%%%%
%\begin{align}
%\label{eq:desnity matrix}
%    \rho(r) = \sum_{\mu}\sum_{\nu}\gamma_{\mu\nu} \bra{\nu}\ket{\mu}, 
%\end{align}
%%%%%%%%%%%%%%%%%%%%%%%%%%%%%%%%%%%%%%%%%%%%%%%%%%%%%%%%%%%%%%%%%%
%%%%%%%%%%%%%%%%%%  End of Eq. 13  %%%%%%%%%%%%%%%%%%%%%%%%%%%%%%%
%%%%%%%%%%%%%%%%%%%%%%%%%%%%%%%%%%%%%%%%%%%%%%%%%%%%%%%%%%%%%%%%%%
After introducing an atomic orbital (AO) basis set, one $\alpha$-component of the dipole moment is evaluated from
%%%%%%%%%%%%%%%%%%%%%%%%%%%%%%%%%%%%%%%%%%%%%%%%%%%%%%%%%%%%%%%%%%
%%%%%%%%%%%%%%%%%%  Eq. 14 (u_contract) %%%%%%%%%%%%%%%%%%%%%%%%%%
%%%%%%%%%%%%%%%%%%%%%%%%%%%%%%%%%%%%%%%%%%%%%%%%%%%%%%%%%%%%%%%%%%
\begin{align}
\label{eq: dipole_qc_def}
\mu_{\alpha} = \sum_{i=1}^{N_{\rm nuc}}Z_{i}R_{i\alpha} - \sum_{\mu}\sum_{\nu}\gamma_\nu^\mu \bra{\nu}\hat{r}_{\alpha}\ket{\mu},
\end{align} 
%%%%%%%%%%%%%%%%%%%%%%%%%%%%%%%%%%%%%%%%%%%%%%%%%%%%%%%%%%%%%%%%%%
%%%%%%%%%%%%%%%%%%  End of Eq. 14  %%%%%%%%%%%%%%%%%%%%%%%%%%%%%%%
%%%%%%%%%%%%%%%%%%%%%%%%%%%%%%%%%%%%%%%%%%%%%%%%%%%%%%%%%%%%%%%%%%
where $\gamma_\nu^\mu$ is the density matrix in the AO basis, and $\bra{\nu}\hat{r}_{\alpha}\ket{\mu}= \int\chi^{*}_{\nu}(r)~r_{\alpha}~\chi_{\mu}(r)dr $ are the dipole moment integrals expressed in the AO basis $\{ \chi_\nu \}$.~\cite{helgaker_book}
Since all pCCD-based methods work in the molecular orbital (MO) basis and hence the corresponding 1-RDMs are defined for the molecular orbitals, we need to perform an AO-MO transformation step of the dipole moment integrals or the 1-RDMs, respectively.

%%%%%%%%%%%%%%%%%%%%%%%%%%%%%%%%%%%%%%%%%%%%%%%%%%%%%%%%%%%%%%%%%%
%%%%%%%%%%%%  Comput. Details          %%%%%%%%%%%%%%%%%%%%%%%%%%%
%%%%%%%%%%%%%%%%%%%%%%%%%%%%%%%%%%%%%%%%%%%%%%%%%%%%%%%%%%%%%%%%%% 
\section{Computational details}\label{sec:computatioinal-details}
%%%%%%%%%%%%%%%%%%%%%%%%%%%%%%%%%%%%%%%%%%%%%%%%%%%%%%%%%%%%%%%%%%
%%%%%%%%%%%%  Comp. Det: Geometries     %%%%%%%%%%%%%%%%%%%%%%%%%%%
%%%%%%%%%%%%%%%%%%%%%%%%%%%%%%%%%%%%%%%%%%%%%%%%%%%%%%%%%%%%%%%%%% 
\subsection{Structures}
%%%%%%%%%%%%%%%%%%%%%%%%%%%%%%%%%%%%%%%%%%%%%%%%%%%%%%%%%%%%%%%%%%
The geometries of the main group diatomic molecules were taken from Liu et al.~\cite{liu2020data} and references therein.
Their bond lengths are collected in Table S1 of the SI.
Each diatomic molecule is placed along the z-axis.

\begin{figure}[h!]
\includegraphics[width=\columnwidth]{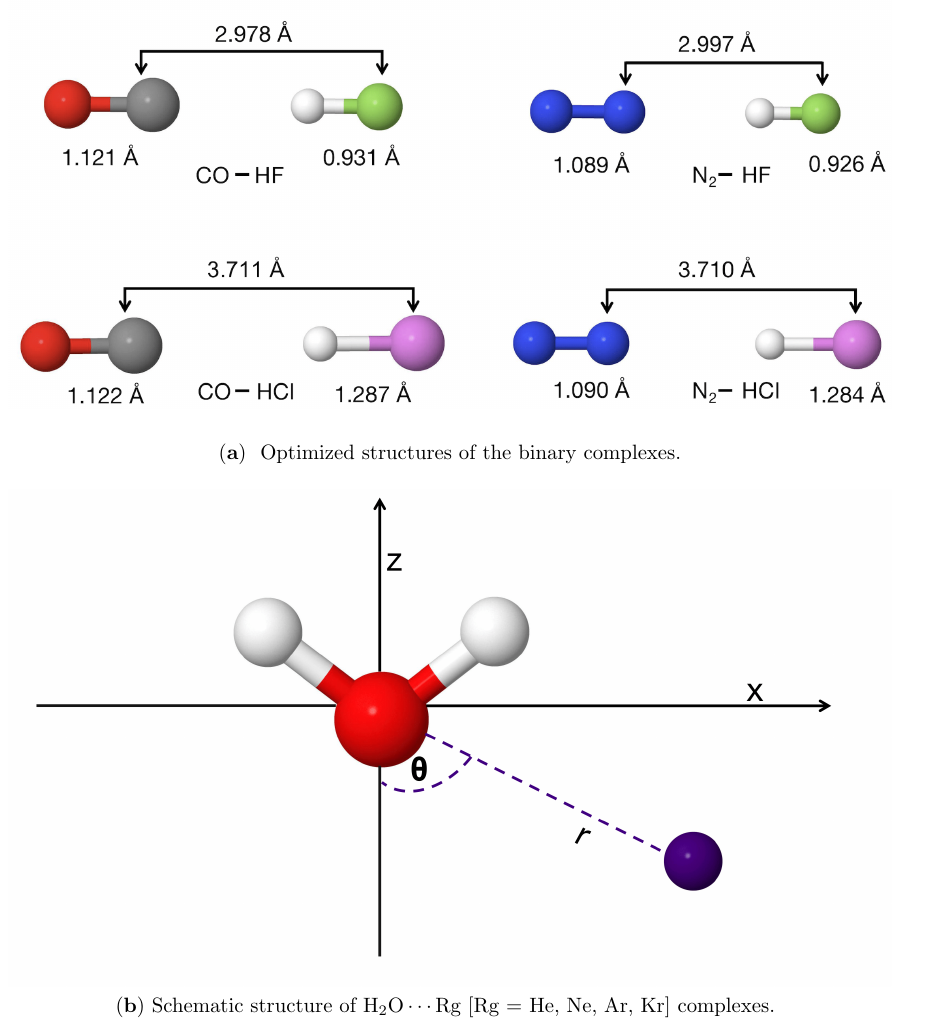}
\caption{Structural representations of the complexes studied in this work.}
\label{fig:str_complex}
\end{figure}

The structures of the \ce{CO-HF}, \ce{CO-HCl}, \ce{N2-HF}, and \ce{N2-HCl}~\cite{curtiss1985investigation, soper1981microwave, cummins1987ab} were optimized with the CCSD(T) method and the augmented Dunning-type correlation consistent basis sets of quadruple-$\zeta$ quality (aug-cc-pVQZ).~\cite{basis_dunning,petersonJCP02_117_10548}
The molecules were placed along the z-axis, as shown in Figure~\ref{fig:str_complex}a along with the optimized bond lengths. 
%%%%%%%%%%%%%%%%%%%%%%%%%%%%%%%%%%%%%%%%%%%%%%%%%%%%%%%%%%%%%%%%%%
%%%%%%%%%%%%%%%%% FIGURE 1 %%%%%%%%%%%%%%%%%%%%%%%%%%%%%%%%%%%%%%%
%%%%%%%%%%%%%%%%%%%%%%%%%%%%%%%%%%%%%%%%%%%%%%%%%%%%%%%%%%%%%%%%%%
%%%%%%%%%%%%%%%%%%%%%%%%%%%%%%%%%%%%%%%%%%%%%%%%%%%%%%%%%%%%%%%%%%
%%%%%%%%%%%%%%%%% END OF FIGURE 1 %%%%%%%%%%%%%%%%%%%%%%%%%%%%%%%%
%%%%%%%%%%%%%%%%%%%%%%%%%%%%%%%%%%%%%%%%%%%%%%%%%%%%%%%%%%%%%%%%%%

The bond parameters of the \ce{H2O\cdots Rg} [Rg = He, Ne, Ar, Kr] complexes were taken from Haskopoulos et al.~\cite{haskopoulos2010interaction} 
Following the original work, these complexes were kept in the xz plane with the center of mass of \ce{H2O} at the origin and the oxygen atom on the negative z-axis (see Figure~\ref{fig:str_complex}b). 
The equilibrium bond parameters of these 4 complexes are given in Table S8 of the SI. 

\subsection{pCCD-based dipole moment} 
%%%%%%%%%%%%%%%%%%%%%%%%%%%%%%%%%%%%%%%%%%%%%%%%%%%%%%%%%%%%%%%%%%
The pCCD-based dipole moment calculations were carried out in a developer version (v1.4.0dev) of the \textsc{PyBEST} software package.~\cite{pybest-paper, pybest-paper-update1-cpc-2024, pybest-gpu-jctc-2024}
The dipole moments were calculated with the Dunning family of basis sets with and without augmentation, that is, (aug-)cc-pVnZ, for n=D, T and Q  with optimized general contractions.~\cite{basis_dunning,deyonkerJPCA07_111_11383,petersonJCP02_117_10548,hillJCP17_147_244106}
Henceforth, the orbital optimized pCCD and the LCC corrections on top of it are called oo-pCCD and oo-pCCD-LCC(S)D. Consequently, the pCCD and pCCD-LCC(S)D will refer to pCCD and a posteriori LCC corrections within a canonical (Hartree--Fock (HF)) orbital basis. 

Cholesky decomposition with a threshold of 10$^{-5}$ was used for all systems. 
Pipek--Mezey orbital localization~\cite{pipek-mezey-localization}  was used to speed up the orbital optimization process for all systems. 
In all pCCD and oo-pCCD based calculations, all non-valence orbitals were kept frozen to match the MOLPRO reference results (vide infra).  
%%%%%%%%%%%%%%%%%%%%%%%%%%%%%%%%%%%%%%%%%%%%%%%%%%%%%%%%%%%%%%%%%%
%%%%%%%%%%%%  Comp. Det: Pybest : emb  %%%%%%%%%%%%%%%%%%%%%%%%%%%
%%%%%%%%%%%%%%%%%%%%%%%%%%%%%%%%%%%%%%%%%%%%%%%%%%%%%%%%%%%%%%%%%% 
\subsubsection{pCCD-in-DFT}
%%%%%%%%%%%%%%%%%%%%%%%%%%%%%%%%%%%%%%%%%%%%%%%%%%%%%%%%%%%%%%%%%%
The embedding potentials were generated within the Amsterdam Modeling Suite (AMS2022)~\cite{adf1, ams2022, adf2}
and then extracted with the help of the PyADF~\cite{pyadf} scripting framework. 
In all DFT-in-DFT calculations, the triple-$\zeta$ double polarization (TZ2P) basis set,~\cite{adf_b} the PW91~\cite{pw91_xc, pbex} exchange--correlation functional, and the PW91k~\cite{pw91k} kinetic energy functional were used.
More details about the DFT-in-DFT frozen density embedding (FDE) setup used here to obtain the embedding potential are described in our previous work.~\cite{pccd-static-embedding} 
For each embedding calculation, two sets of calculations were performed, in which the system and environment were swapped, and their dipole moment results were added together.  
%%%%%%%%%%%%%%%%%%%%%%%%%%%%%%%%%%%%%%%%%%%%%%%%%%%%%%%%%%%%%%%%%%
%%%%%%%%%%%%  Comp. Det: reference     %%%%%%%%%%%%%%%%%%%%%%%%%%%
%%%%%%%%%%%%%%%%%%%%%%%%%%%%%%%%%%%%%%%%%%%%%%%%%%%%%%%%%%%%%%%%%% 
\subsection{Reference dipole moment calculations}
%%%%%%%%%%%%%%%%%%%%%%%%%%%%%%%%%%%%%%%%%%%%%%%%%%%%%%%%%%%%%%%%%% 
All reference values were obtained using the MOLPRO package version~19.~\cite{molpro-wires, molpro2020_jcp, molpro2020-authors} 
The reference dipole moments were obtained using the CCSD and CCSD(T) methods (relaxed and unrelaxed density matrices) and the same family of basis sets used in pCCD and oo-pCCD based calculations with PyBEST.
In this work, CCSD$_u$ and CCSD(T)$_u$ refer to dipole moments with unrelaxed densities whereas CCSD$_r$ and CCSD(T)$_r$ are for the same with relaxed densities. 

%%%%%%%%%%%%%%%%%%%%%%%%%%%%%%%%%%%%%%%%%%%%%%%%%%%%%%%%%%%%%%%%%%
%%%%%%%%%%%%  Results:                 %%%%%%%%%%%%%%%%%%%%%%%%%%%
%%%%%%%%%%%%%%%%%%%%%%%%%%%%%%%%%%%%%%%%%%%%%%%%%%%%%%%%%%%%%%%%%% 
%%%%%%%%%%%%%%%%%%%%%%%%%%%%%%%%%%%%%%%%%%%%%%%%%%%%%%%%%%%%%%%%%% 
%%%%%%%%%%%%%%%%%%%%%%%%%%%%%%%%%%%%%%%%%%%%%%%%%%%%%%%%%%%%%%%%%% 
\section{Results and Discussion}\label{sec:results}
%%%%%%%%%%%%%%%%%%%%%%%%%%%%%%%%%%%%%%%%%%%%%%%%%%%%%%%%%%%%%%%%%% 
%%%%%%%%%%%%%%%%%%%%%%%%%%%%%%%%%%%%%%%%%%%%%%%%%%%%%%%%%%%%%%%%%% 
%%%%%%%%%%%%%%%%%%%%%%%%%%%%%%%%%%%%%%%%%%%%%%%%%%%%%%%%%%%%%%%%%% 
\subsection{Dipole moment of main group diatomics}\label{sec:results-diatomics}
%%%%%%%%%%%%%%%%%%%%%%%%%%%%%%%%%%%%%%%%%%%%%%%%%%%%%%%%%%%%%%%%%% 
%%%%%%%%%%%%%%%%%%%%%%%%%%%%%%%%%%%%%%%%%%%%%%%%%%%%%%%%%%%%%%%%%% 

%%%%%%%%%%%%%%%%%%%%%%%%%%%%%%%%%%%%%%%%%%%%%%%%%%%%%%%%%%%%%%%%%% 
%%%%%%%%%%%%%%%%%%%%%%%%%%%%%%%%%%%%%%%%%%%%%%%%%%%%%%%%%%%%%%%%%% 
\subsubsection{Statistical analysis}
%%%%%%%%%%%%%%%%%%%%%%%%%%%%%%%%%%%%%%%%%%%%%%%%%%%%%%%%%%%%%%%%%% 
%%%%%%%%%%%%%%%%%%%%%%%%%%%%%%%%%%%%%%%%%%%%%%%%%%%%%%%%%%%%%%%%%% 
We start our analysis with the diatomic molecules and the basis set dependence.
Table S2 of the SI collects all the dipole moments computed with different quantum chemistry methods and (aug-)cc-pVnZ [n=D, T, Q] basis sets with and without augmented functions. 
All basis sets provide qualitatively similar results.
The most significant differences are observed between the cc-pVDZ and cc-pVTZ basis sets, and between the standard and augmented series.
The differences within the augmented series are significantly smaller.
Table S3 collects the mean unsigned errors (MUE) and root-mean-square errors (RMSE) for all the methods considered in this work in all basis sets with respect to experimental dipole moments.
We observe that triple-$\zeta$ and quadruple-$\zeta$ basis sets produce similar errors. 
MUE and RMSE increase slightly from aug-cc-pVTZ to aug-cc-pVQZ for oo-pCCD and oo-pCCD-LCCD. However, the opposite is seen for oo-pCCD-LCCSD.
In short, not much accuracy is gained by increasing the size of the basis set from triple-$\zeta$ to quadruple-$\zeta$ in terms of dipole moments, as has been observed in previous works with traditional coupled cluster methods.~\cite{hickey2014benchmarking}
To that end, we used the aug-cc-pVTZ as the basis set of choice for further investigations. 
In addition, we should stress that the dipole moment results are more or less independent of the frozen core approximation (cf. Table S4 of the SI).  

%%%%%%%%%%%%%%%%%%%%%%%%%%%%%%%%%%%%%%%%%%%%%%%%%%%%%%%%%%%%%%%%%% 
%%%%%%%%%%%%%%%%%%%    TABLE  1  %%%%%%%%%%%%%%%%%%%%%%%%%%%%%%%%%
%%%%%%%%%%%%%%%%%%%%%%%%%%%%%%%%%%%%%%%%%%%%%%%%%%%%%%%%%%%%%%%%%% 
\newcolumntype{P}[1]{>{\centering\arraybackslash}p{#1}}
\begin{table*}[ht!]
%\setlength{\tabcolsep}{2.3pt}
%\begin{footnotesize}
\caption{Error analysis for the dataset of 20 main group diatomics studied in this work. Errors are calculated in Debye using the aug-cc-pVTZ basis for all methods. MUE and RMSE stand for mean unsigned error [$\frac{1}{N}\sum_{i}^{N}\lvert\mu_{\rm Method,i} - \mu_{\rm ref.,i}\rvert$] and root mean square error [$\sqrt{(\sum_{i}^{N}(\mu_{\rm Method,i} - \mu_{\rm ref.,i})^2)/N)}$~], respectively, where $N$ is the number of molecules in the dataset.
For CCSD(T)$_r$ reference data, MUE and RMSE are divided for the full data set, all singly-bonded, and multiply-bonded systems (with and without the outlier MgO).\\}
\label{tbl:diatomics-statistic}
\begin{tabular*}{\textwidth}{lccccccccccccccccccc}
\hline
\centering
\multirow{2}{*}{Method} &\multicolumn{2}{c}{Exp.} & &\multicolumn{2}{c}{CCSD(T)$_{u}$} & &\multicolumn{11}{c}{CCSD(T)$_r$} \\
 && & & && &\multicolumn{2}{c}{full data set} & &\multicolumn{2}{c}{singly-bonded} & &\multicolumn{2}{c}{multiply-bonded}& &\multicolumn{2}{c}{w/o MgO}\\
 \cline{2-3}\cline{5-6}\cline{8-9}\cline{11-12}\cline{14-15}\cline{17-18}
              &MUE   &RMSE &  &MUE   &RMSE & &MUE   &RMSE & &MUE   &RMSE & &MUE   &RMSE & &MUE   &RMSE\\
              \hline
pCCD            & 0.437 & 0.640  &&~0.471 &~0.782 && 0.393 & 0.585    && 0.131 &  0.155 && 0.577  & 0.807 && 0.437 &  0.512  \\
pCCD-LCCD       & 0.356 & 0.535  &&~0.365 &~0.663 && 0.288 & 0.460    && 0.091 &  0.109 && 0.429  & 0.638 && 0.309 &  0.374  \\
pCCD-LCCSD      & 0.530 & 0.753  &&~0.382 &~0.547 && 0.465 & 0.730    && 0.105 &  0.138 && 0.754  & 1.024 && 0.585 &  0.672  \\
CCSD$_{u}$      & 0.180 & 0.276  &&~0.136 &~0.307 && 0.063 & 0.102    && 0.032 &  0.061 && 0.086  & 0.131 && 0.059 &  0.068  \\
CCSD(T)$_{u}$   & 0.194 & 0.300  &&---    &---    && 0.087 & 0.217    && 0.011 &  0.016 && 0.149  & 0.306 && 0.070 &  0.081  \\
oo-pCCD         & 0.363 & 0.604  &&~0.323 &~0.525 && 0.336 & 0.637    && 0.078 &  0.097 && 0.550  & 0.896 && 0.375 &  0.526  \\
oo-pCCD-LCCD    & 0.345 & 0.581  &&~0.284 &~0.474 && 0.302 & 0.605    && 0.068 &  0.081 && 0.493  & 0.852 && 0.309 &  0.429  \\
oo-pCCD-LCCSD   & 0.373 & 0.476  &&~0.248 &~0.307 && 0.283 & 0.352    && 0.092 &  0.136 && 0.399  & 0.465 && 0.393 &  0.441  \\
CCSD$_{r}$      & 0.191 & 0.262  &&~0.169 &~0.289 && 0.085 &~0.119    && 0.038 &  0.062 && 0.122  & 0.156 && 0.116 &  0.144  \\
\hline
\end{tabular*}
%\end{footnotesize}
\end{table*}
%%%%%%%%%%%%%%%%%%%%%%%%%%%%%%%%%%%%%%%%%%%%%%%%%%%%%%%%%%%%%%%%%% 
%%%%%%%%%%%%%%%%%%%  END OF TABLE  1  %%%%%%%%%%%%%%%%%%%%%%%%%%%%
%%%%%%%%%%%%%%%%%%%%%%%%%%%%%%%%%%%%%%%%%%%%%%%%%%%%%%%%%%%%%%%%%% 
Table~\ref{tbl:diatomics-statistic} summarizes the MUE and the RMSE of our pCCD-based methods with respect to the experimental data and the reference theoretical CCSD(T)$_{r}$ and CCSD(T)$_{u}$ values.
The data from Table~\ref{tbl:diatomics-statistic} shows that, on average, the orbital optimization within the pCCD reference function improves the overall performance of the pCCD-based dipole moments with respect to experiment and reference theoretical data. 
Including LCC on top of pCCD further refines the dipole moment values towards the reference. 
From a numerical perspective, the MUEs for pCCD and pCCD-LCCSD improve by $\approx$0.1 D upon the addition of orbital optimization.
However, pCCD-LCCD statistics do not show much improvement with the same. 

\begin{figure}[h!]
   \includegraphics[width=\columnwidth]{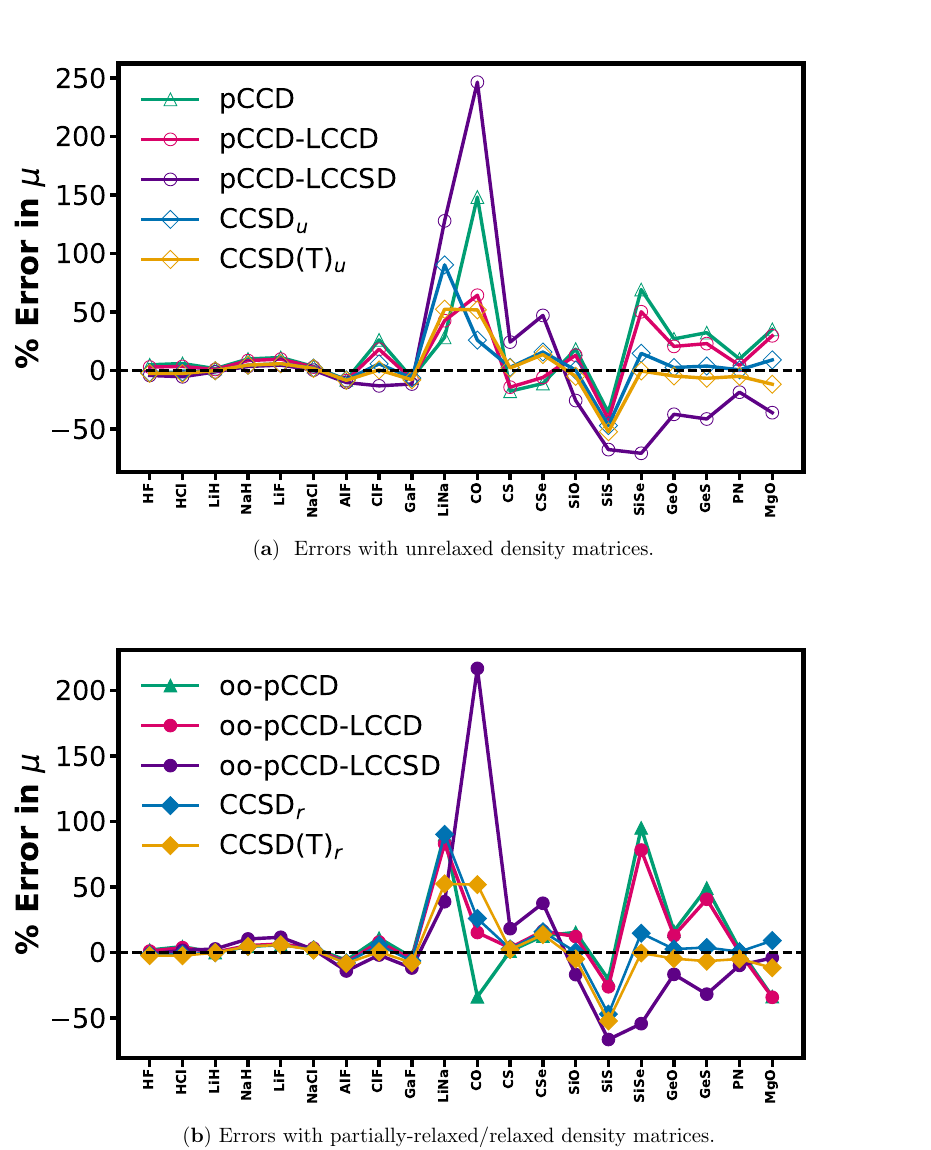} 
   \caption{Percentage errors in all methods using  aug-cc-pVTZ basis  with respect to the experimental dipole moment values for all molecules in the data set.}
   \label{fig:error_plot}
\end{figure}

\begin{figure}[h!]
\centering
\includegraphics[width=\columnwidth]{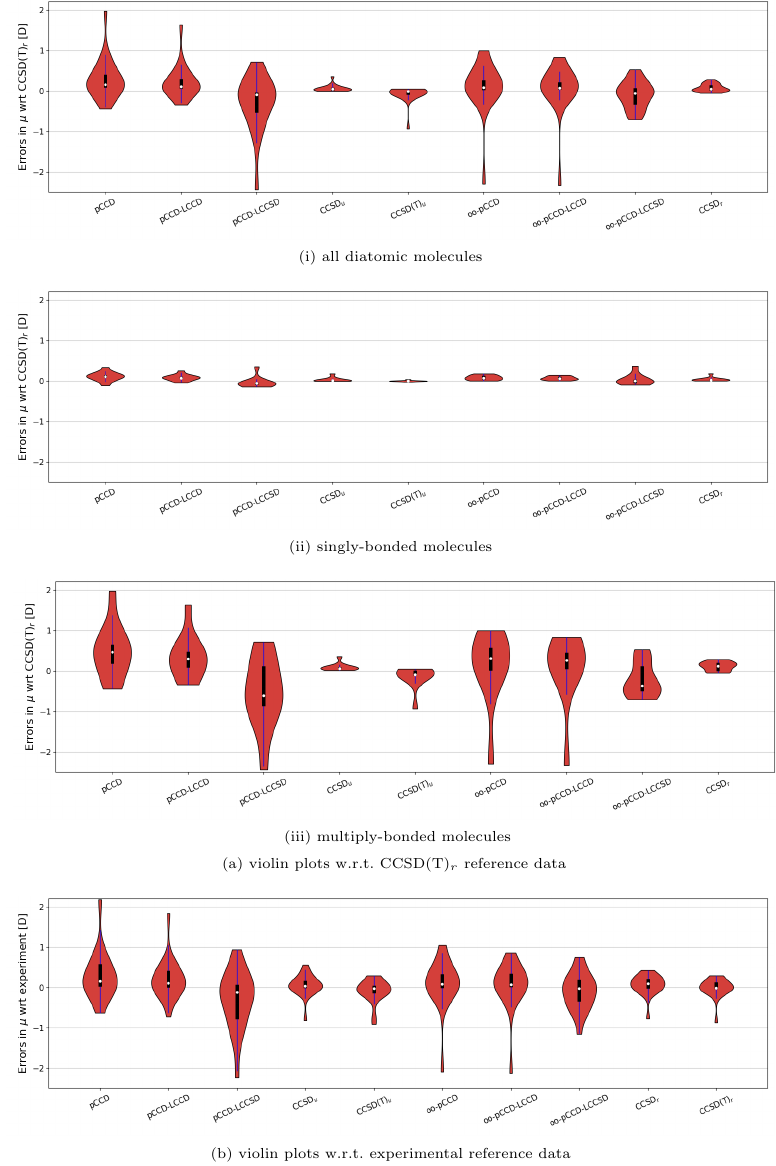}
\caption{Violin plots illustrating errors (in D) derived from selected methods (refer to Table S2 for numerical values). All errors are reported relative to either (a) CCSD(T)$_{r}$ or (b) experimental reference data. A dot in each violin plot represents the median value, while the blue line indicates the 1.5 interquartile range and the black bar the quartile range, respectively.}
\label{fig:violin-plots}
\end{figure}

In Figure~\ref{fig:error_plot}, we show the percentage errors (with sign) in dipole moments obtained with pCCD-based methods for individual molecules, with respect to the experimental values. 
Figure~\ref{fig:error_plot}a shows the performance of pCCD and its variants without orbital optimization, i.e., with completely unrelaxed densities, whereas Figure ~\ref{fig:error_plot}b depicts the same for oo-pCCD and subsequent LCC variants, with relaxed densities achieved through orbital optimization within pCCD. 
Here, it is important to remember that the oo-pCCD-LCC density matrices are only partially relaxed. 
In this plot, we see a clear distinction between the behavior of simple 
singly-bonded molecules and the molecules with significant multiple-bond characters. 
As evident from Figure~\ref{fig:error_plot}b, the second class of molecules shows higher relative errors with all pCCD-based methods. 
We also observe that LCCD values remain in close vicinities of the pCCD ones for most of the molecules. 
Exceptions to this occur for molecules, again, with multiple bond characters (see also last columns in Table~\ref{tbl:diatomics-statistic}).
The LCCSD values, on the other hand, differ significantly from their counterparts for almost all molecules. 
Of particular interest is the MgO molecule, where the oo-pCCD-LCCSD seems to perform even better than CCSD(T)$_r$ with respect to the experiment.
The impact of the character of the bond on the dipole moment values obtained with pCCD-based methods is also evident in the violin plots in Figure~\ref{fig:violin-plots}. 
Specifically, Figure~\ref{fig:violin-plots}a and Figure~\ref{fig:violin-plots}b show the distribution (skewness) of the errors in dipole moments with pCCD-based methods with respect to CCSD(T)$_r$ and experimental values, respectively.
As can be seen, the multiply-bonded molecules show a significantly higher spread of errors than the singly-bonded molecules.
For the latter, the interquartile ranges are distributed closely around the median.
If the orbitals are optimized within pCCD, the median and spread are shifted closer to the reference.
Moreover, an LCCSD correction introduces outliers and features a broader interquartile range.
For multiply-bonded systems, the skewness of errors is right-shifted for (oo-)pCCD and (oo)-pCCD-LCCD, while (oo-)-pCCD-LCCSD yields left-shifted ones.
Furthermore, (oo)-pCCD-LCCD reduces the interquartile range and shifts the median closer to the reference, while (oo)-pCCD-LCCSD introduces a strong asymmetry, moving the median below the reference point.

Overall, though our statistical analysis shows the utility of adding dynamic correlation with LCC corrections in the pCCD framework, a case-by-case analysis reveals that this is not a black-box tool for all molecules regarding the calculation of dipole moments.
That motivates us to conduct a deeper analysis of the performance of pCCD-based methods for different types of molecules and bonding patterns in the next section.

%%%%%%%%%%%%%%%%%%%%%%%%%%%%%%%%%%%%%%%%%%%%%%%%%%%%%%%%%%%%%%%%% 
%%%%%%%%%%%%%%%%%%%    Figure  2  %%%%%%%%%%%%%%%%%%%%%%%%%%%%%%%%%
%%%%%%%%%%%%%%%%%%%%%%%%%%%%%%%%%%%%%%%%%%%%%%%%%%%%%%%%%%%%%%%%%% 
%%%%%%%%%%%%%%%%%%%%%%%%%%%%%%%%%%%%%%%%%%%%%%%%%%%%%%%%%%%%%%%%%% 
%%%%%%%%%%%%%%%%%%%    Figure  2 Ends  %%%%%%%%%%%%%%%%%%%%%%%%%%%
%%%%%%%%%%%%%%%%%%%%%%%%%%%%%%%%%%%%%%%%%%%%%%%%%%%%%%%%%%%%%%%%%% 

%%%%%%%%%%%%%%%%%%%%%%%%%%%%%%%%%%%%%%%%%%%%%%%%%%%%%%%%%%%%%%%%%% 
%%%%%%%%%%%%%%%%%%%    Figure  3  %%%%%%%%%%%%%%%%%%%%%%%%%%%%%%%%%
%%%%%%%%%%%%%%%%%%%%%%%%%%%%%%%%%%%%%%%%%%%%%%%%%%%%%%%%%%%%%%%%%% 
%%%%%%%%%%%%%%%%%%%%%%%%%%%%%%%%%%%%%%%%%%%%%%%%%%%%%%%%%%%%%%%%%% 
%%%%%%%%%%%%%%%%%%%    Figure  3 Ends  %%%%%%%%%%%%%%%%%%%%%%%%%%%
%%%%%%%%%%%%%%%%%%%%%%%%%%%%%%%%%%%%%%%%%%%%%%%%%%%%%%%%%%%%%%%%%% 

%%%%%%%%%%%%%%%%%%%%%%%%%%%%%%%%%%%%%%%%%%%%%%%%%%%%%%%%%%%%%%%%%% 
%%%%%%%%%%%%%%%%%%%%%%%%%%%%%%%%%%%%%%%%%%%%%%%%%%%%%%%%%%%%%%%%%% 
\subsubsection{In-depth comparison with reference theoretical methods}
\begin{figure*}
   \includegraphics[width=0.8\textwidth]{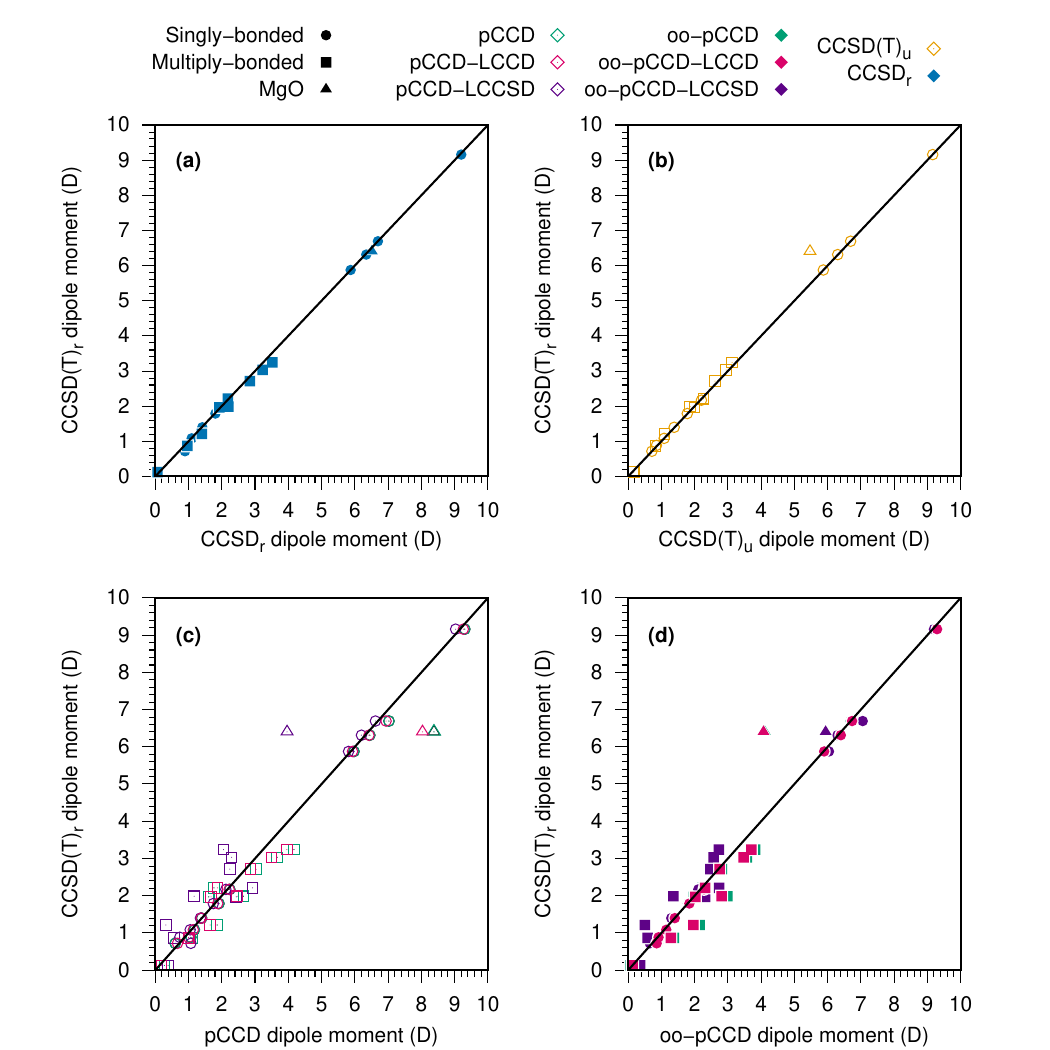} 
   \caption{The correlation between the reference CCSD(T)$_{r}$ dipole moments (in D) and other CC-based methods.
   \textbf{(a)} relaxed CCSD; \textbf{(b)} unrelaxed CCSD(T); \textbf{(c)} pCCD and pCCD with LCC corrections; and \textbf{(d)} oo-pCCD and oo-pCCD with LCC corrections.
   }
   \label{fig:comp_all}
\end{figure*}

%%%%%%%%%%%%%%%%%%%%%%%%%%%%%%%%%%%%%%%%%%%%%%%%%%%%%%%%%%%%%%%%%% 
%%%%%%%%%%%%%%%%%%%%%%%%%%%%%%%%%%%%%%%%%%%%%%%%%%%%%%%%%%%%%%%%%% 
Figure~\ref{fig:comp_all}a shows the correlation between the reference CCSD(T)$_{r}$ and the CCSD$_{r}$ dipole moments (both with relaxed density matrices).
We observe an excellent agreement between the two methods for singly-bonded molecules (represented by circles). 
The correlation worsens for multiply-bonded systems (marked by squares), underlining the importance of triple excitations.
Figure~\ref{fig:comp_all}b shows good agreement between CCSD(T) results using relaxed and unrelaxed density matrices.
The only exception is the MgO molecule (denoted by a triangular shape in Figure~\ref{fig:comp_all}), for which relaxation has a more profound effect.
%%%%%%%%%%%%%%%%%%%%%%%%%%%%%%%%%%%%%%%%%%%%%%%%%%%%%%%%%%%%%%%%%%
%%%%%%%%%%%%%%%%% FIGURE 4 %%%%%%%%%%%%%%%%%%%%%%%%%%%%%%%%%%%%%%%
%%%%%%%%%%%%%%%%%%%%%%%%%%%%%%%%%%%%%%%%%%%%%%%%%%%%%%%%%%%%%%%%%%
%%%%%%%%%%%%%%%%%%%%%%%%%%%%%%%%%%%%%%%%%%%%%%%%%%%%%%%%%%%%%%%%%%
%%%%%%%%%%%%%%%%% END OF FIGURE 4 %%%%%%%%%%%%%%%%%%%%%%%%%%%%%%%%
%%%%%%%%%%%%%%%%%%%%%%%%%%%%%%%%%%%%%%%%%%%%%%%%%%%%%%%%%%%%%%%%%%
By comparing the pCCD-based dipole moments with CCSD(T)$_{r}$, we observe a set of characteristic features for each molecule type.
Molecules with negligible relaxation effects and triple excitations dependence (mainly singly-bonded) provide a very satisfactory agreement between all pCCD-based methods and reference results (cf. Figure~\ref{fig:comp_all}c-d).
Although the variation among pCCD-based methods is slight, we note that the pCCD-LCCSD variant using the canonical orbitals leads to the smallest errors.
On the contrary, when orbital-optimized pCCD orbitals are employed, the LCCD correction is the most reliable and results in the smallest errors. 
Surprisingly, the LCCSD correction on top of oo-pCCD increases the error in some cases.

The MgO molecule presents the most challenging test case for pCCD-LCCD and pCCD-LCCSD methods (cf. Figure~\ref{fig:comp_all}c).
The oo-pCCD-LCCD dipole moment is similar to the pCCD-LCCSD using canonical HF orbitals, which suggests that the orbital relaxation has recovered the effect of the linearized single excitations (compare Figures~\ref{fig:comp_all}c and ~\ref{fig:comp_all}d). 
With the LCCSD correction on top of the oo-pCCD, the dipole moment agreement with the CCSD(T)$_{\rm r}$ reference value improves significantly.
Specifically, the absolute (and relative) error in the MgO dipole moment reduces from 2.23~D (36\%) to 0.46~D (7\%) when moving from oo-pCCD-LCCD to oo-pCCD-LCCSD, respectively. 

The diatomic molecules with a large contribution of triple excitations to the dipole moment show a similar, but smaller, swing in dipole values between the pCCD-LCCD and pCCD-LCCSD as the one seen for the MgO molecule.
However, as the main change in dipole moments is not due to an orbital relaxation effect, the oo-pCCD variation leads to a dipole value closer to the pCCD than the pCCD-LCCSD one.
Consequently, the oo-pCCD-LCCD and oo-pCCD-LCCSD results approach the reference from opposite directions.
Although the orbital optimization improves the results, the oo-pCCD-LCCD and oo-pCCD-LCCSD dipole moment values have similar but substantial errors.
The only exception is the carbon-containing compounds; in these cases, the oo-pCCD-LCCSD error to the CCSD(T)$_{r}$ is higher than the oo-pCCD-LCCD.
Once some of the studied systems require triple excitations, as concluded during the analysis of Figure~\ref{fig:comp_all}a, none of the investigated pCCD-based approaches can recover this effect, and, therefore, such an error is expected.

Based on this analysis, the variation of dipole moment values among pCCD, oo-pCCD, and pCCD-LCCSD results can be used to estimate the magnitude of the orbital relaxation and triple excitations for the dipole moment.
Systems, where the three values agree with each other have a small dependence on orbital relaxation and triple excitations.
Thus, either the pCCD-LCCSD or oo-pCCD-LCCD leads to minor errors with respect to the CCSD(T)$_{r}$ reference, that is, a relative average error of around 4\%.
When oo-pCCD and pCCD-LCCSD are similar, orbital relaxation is required, and the oo-pCCD-LCCSD value should be preferable.
Lastly, for distinct oo-pCCD and pCCD-LCCSD values, pCCD-based methods would require a larger excitation order to be reliable.
In these cases, excluding the carbon-containing molecules, both oo-pCCD-LCCD and oo-pCCD-LCCSD methods have a relative average error of around 30\%.
Including the carbon-based ones, the oo-pCCD-LCCD error decreases to 21\%, while the oo-pCCD-LCCSD one increases up to 43\%.
%%%%%%%%%%%%%%%%%%%%%%%%%%%%%%%%%%%%%%%%%%%%%%%%%%%%%%%%%%%%%%%%%%
%%%%%%%%%%%%%%%%%%%%%%%%%%%%%%%%%%%%%%%%%%%%%%%%%%%%%%%%%%%%%%%%%%
\subsection{Dipole Moment Surfaces with pCCD-based Methods}
%%%%%%%%%%%%%%%%%%%%%%%%%%%%%%%%%%%%%%%%%%%%%%%%%%%%%%%%%%%%%%%%%%
%%%%%%%%%%%%%%%%%%%%%%%%%%%%%%%%%%%%%%%%%%%%%%%%%%%%%%%%%%%%%%%%%%
Dipole moment surfaces (DMS) are essential for estimating rovibrational spectroscopic parameters of molecules. 
Here, we focus on the DMS of two main group diatomic molecules, HF and CO. 
Their DMSs have been widely studied~\cite{buldakov2009_dipole-sign, lykhin2021dipole} in previous theoretical works and, thus, represent suitable test cases for the investigated pCCD-based methods in different bond length regions. 
In this work, the diatomics AB are placed along the z-axis with A (the less electronegative atom) at the origin and B on the positive z-axis. 
Then, the bond between the two atoms of AB is stretched along the positive z-axis for constructing the DMS. 
Hence, a positive $\mu_z$ value will indicate A$^-$B$^+$ polarity, whereas a negative $\mu_z$ indicates the same as A$^+$B$^-$. 
%%%%%%%%%%%%%%%%%%%%%%%%%%%%%%%%%%%%%%%%%%%%%%%%%%%%%%%%%%%%%%%%%%
%%%%%%%%%%%%%%%%%%%%%%%%%%%%%%%%%%%%%%%%%%%%%%%%%%%%%%%%%%%%%%%%%%
\subsubsection{Hydrogen Fluoride (HF)}\label{sec:hf}
%%%%%%%%%%%%%%%%%%%%%%%%%%%%%%%%%%%%%%%%%%%%%%%%%%%%%%%%%%%%%%%%%%
%%%%%%%%%%%%%%%%%%%%%%%%%%%%%%%%%%%%%%%%%%%%%%%%%%%%%%%%%%%%%%%%%%

%%%%%%%%%%%%%%%%%%%%%%%%%%%%%%%%%%%%%%%%%%%%%%%%%%%%%%%%%%%%%%%%%%
%%%%%%%%%%%%  FIGURE                  %%%%%%%%%%%%%%%%%%%%%%%%%%%
%%%%%%%%%%%%%%%%%%%%%%%%%%%%%%%%%%%%%%%%%%%%%%%%%%%%%%%%%%%%%%%%%% 
\begin{figure}[h!]
    \centering
    \includegraphics[width=\columnwidth]{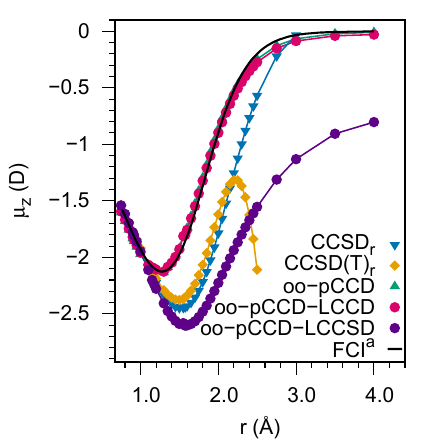}  
    \caption{Dipole moment surface of HF in aug-cc-pVTZ basis. $^a$FCI/cc-pVDZ DMS is taken from Samanta and K{\"o}hn.~\cite{mrcc-first-order-properties-jcp-2018}}    
    \label{fig:dms_hf}
\end{figure}
%%%%%%%%%%%%%%%%%%%%%%%%%%%%%%%%%%%%%%%%%%%%%%%%%%%%%%%%%%%%%%%%%%
%%%%%%%%%%%%  END OF FIGURE           %%%%%%%%%%%%%%%%%%%%%%%%%%%
%%%%%%%%%%%%%%%%%%%%%%%%%%%%%%%%%%%%%%%%%%%%%%%%%%%%%%%%%%%%%%%%%% 
Figure~\ref{fig:dms_hf} shows the DMS of the HF molecule in the aug-cc-pVTZ basis, calculated with oo-pCCD-based methods.
We also included the CCSD and CCSD(T) DMSs (both with relaxed densities, i.e., CCSD$_{r}$ and CCSD(T)$_{r}$) and the FCI DMS (determined for the cc-pVDZ basis set)~\cite{mrcc-first-order-properties-jcp-2018} for comparison. 
Around the equilibrium distance ($r_{\rm e}=$ 0.917 \AA{}), all oo-pCCD variants agree well with CCSD$_{r}$ and CCSD(T)$_{r}$, as discussed for singly-bonded systems in section~\ref{sec:results-diatomics}.
Passed that region, significant deviations are observed between the curves of oo-pCCD variants and the conventional CC curves. 
Orbital relaxation has become essential in that region. 
The CCSD$_r$ and CCSD(T)$_r$ dipole moment values significantly deviate from the FCI results.
As discussed by Samanta and Köhn,~\cite{mrcc-first-order-properties-jcp-2018} in this region, the CCSD is unable to compensate the ionic contribution of the Hartree-Fock reference wave function.
Although the inclusion of full triple excitations (CCSDT) can improve the CCSD poor modeling, it is not a reasonable zeroth-order wave function for the inclusion of triple excitations perturbatively.
This poor description by CC methods during bond-stretching is reinforced by the change in the DMS behavior beyond 2.00~\AA{} and the lack of convergence of coupled perturbed Hartree--Fock (CPHF) calculations for CCSD(T)$_{r}$ at 2.25~\AA{}.
Therefore, the CCSD(T)$_r$ dipole moment values are not reliable beyond this point. 

The oo-pCCD and oo-pCCD-LCCD DMS lie on top of each other for almost the entire bond length region, indicating the lower significance of the doubles correction on top of pCCD. 
In good agreement with the previous FCI results,~\cite{mrcc-first-order-properties-jcp-2018} both the oo-pCCD and oo-pCCD-LCCD dipole moment curves show turnings at around 1.30-1.35 \AA{} and present a much shallower DMS compared to the other methods from Figure~\ref{fig:dms_hf}.
These results indicate that the oo-pCCD and oo-pCCD-LCCD can model the HF dipole moment at the bond stretching and dissociation regions.
Both have the right shape at larger interatomic distances and converge to the proper asymptotic limit. 
The oo-pCCD-LCCSD curve, on the other hand, overlaps with the CC curves to a slightly longer bond distance.
It also turns at a greater bond length (around 1.60-1.65 \AA{}), showing closer agreement with the turning of CC curves (around 1.50-1.55 \AA{}). 
At stretched bond lengths, the oo-pCCD-LCCSD curve remains below the CC curve and does not converge to the correct asymptotic limit. 
That indicates that the linearized singles correction on top of the oo-pCCD wave function modifies the dipole towards the CCSD results but over-shoots it at stretched geometries. 
%%%%%%%%%%%%%%%%%%%%%%%%%%%%%%%%%%%%%%%%%%%%%%%%%%%%%%%%%%%%%%%%%%
%%%%%%%%%%%%%%%%%%%%%%%%%%%%%%%%%%%%%%%%%%%%%%%%%%%%%%%%%%%%%%%%%%
 \subsubsection{Carbon Monoxide (CO)}\label{sec:co}
\begin{figure}[h!]
    \centering
    \includegraphics[width=\columnwidth]{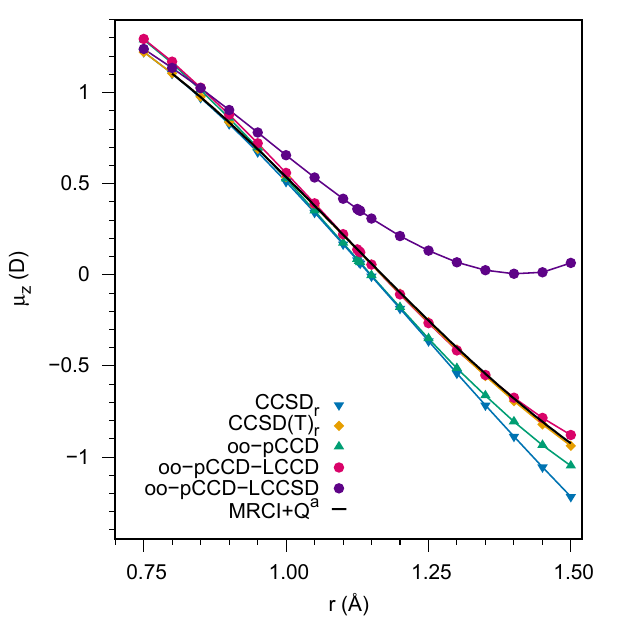}  
    \caption{Dipole moment surface of CO in aug-cc-pVTZ basis. $^a$The MRCI+Q/aug-cc-pCV6Z values have been taken from Balashov et al.~\cite{co-dm-atomowocy-jcp-2023}  }    
    \label{fig:dms_co}
\end{figure}

%%%%%%%%%%%%%%%%%%%%%%%%%%%%%%%%%%%%%%%%%%%%%%%%%%%%%%%%%%%%%%%%%%
%%%%%%%%%%%%%%%%%%%%%%%%%%%%%%%%%%%%%%%%%%%%%%%%%%%%%%%%%%%%%%%%%%
We focused on the region from 0.75 to 1.50 \AA{} in the CO DMS study.
The HF and pCCD wave function optimization beyond that region is very challenging~\cite{ola-tcc} and will likely not provide reliable dipole moments.
In this range of interest of inter-nuclear distances, the CCSD(T)$_{r}$ shows a remarkable agreement with the fitted MRCISD+Q dipole values using the finite-field approach and aug-cc-pCV6D basis set~\cite{co-dm-atomowocy-jcp-2023} as shown in Figure~\ref{fig:dms_co}.
As discussed in section~\ref{sec:results-diatomics}, for the CO case, triple excitations are relevant from the equilibrium distance (around 1.13 \AA{}) onwards. 
That is indicated by the growing splitting between CCSD$_{r}$ and CCSD(T)$_{r}$ dipole values in Figure~\ref{fig:dms_co}.

Similar to what we observed for HF curves, oo-pCCD-LCCSD over-estimates the CO dipole value for large equilibrium distances and has small errors only at the repulsive region (see Figure~\ref{fig:dms_hf}).
To that end, the oo-pCCD-LCCSD DMS of CO is not reliable.
On the other hand, the oo-pCCD DMS matches the CCSD$_{r}$ from 0.90 to 1.25 \AA{} and the oo-pCCD-LCCD DMS resembles the shape of CCSD(T)$_{r}$ up to 1.28 \AA{}.
Throughout the inter-nuclear distances, the average absolute error in the dipole moment of oo-pCCD-LCCD compared to the MRCI+Q reference is about 0.023 D (or around 4\% considering relative errors).
Thus, the oo-pCCD-LCCD provides comparable DMS with the computationally more expensive multireference and CCSD(T)$_{r}$ calculations.

%%%%%%%%%%%%%%%%%%%%%%%%%%%%%%%%%%%%%%%%%%%%%%%%%%%%%%%%%%%%%%%%%%
%%%%%%%%%%%%  FIGURE                  %%%%%%%%%%%%%%%%%%%%%%%%%%%
%%%%%%%%%%%%%%%%%%%%%%%%%%%%%%%%%%%%%%%%%%%%%%%%%%%%%%%%%%%%%%%%%% 
%%%%%%%%%%%%%%%%%%%%%%%%%%%%%%%%%%%%%%%%%%%%%%%%%%%%%%%%%%%%%%%%%%
%%%%%%%%%%%%  END OF FIGURE           %%%%%%%%%%%%%%%%%%%%%%%%%%%
%%%%%%%%%%%%%%%%%%%%%%%%%%%%%%%%%%%%%%%%%%%%%%%%%%%%%%%%%%%%%%%%%% 

%%%%%%%%%%%%%%%%%%%%%%%%%%%%%%%%%%%%%%%%%%%%%%%%%%%%%%%%%%%%%%%%%%
%%%%%%%%%%%%  Results: pCCD-in-DFT     %%%%%%%%%%%%%%%%%%%%%%%%%%%
%%%%%%%%%%%%%%%%%%%%%%%%%%%%%%%%%%%%%%%%%%%%%%%%%%%%%%%%%%%%%%%%%% 
\subsection{Dipole moments from pCCD-based static embedding}\label{sec: emb-results}
%%%%%%%%%%%%%%%%%%%%%%%%%%%%%%%%%%%%%%%%%%%%%%%%%%%%%%%%%%%%%%%%%% 
%%%%%%%%%%%%%%%%%%%%%%%%%%%%%%%%%%%%%%%%%%%%%%%%%%%%%%%%%%%%%%%%%% 
Dipole moments are often used to assess the performance of DFT-based embedding approaches.~\cite{jacob2005orbital} 
The calculated dipole moments are susceptible to electron density changes caused by environmental effects and, thus, are valuable measures for validating the quality of the embedding potential.~\cite{pawel3, fde-nmr-ijqc-2021}
To that end, we investigate the performance of recently implemented pCCD-in-DFT static embedding models~\cite{pccd-static-embedding} for two sets of weakly interacting systems: linear hydrogen-bonded binary complexes and coplanar water complexes with noble gases.
Their structural parameters are presented in Figures~\ref{fig:str_complex}a and~\ref{fig:str_complex}b, respectively.
Building on the experience gained in the previous section and knowing the importance of orbital relaxation in oo-pCCD, we solely focused on orbital-optimized variants.
The supramolecular oo-pCCD-LCCSD dipole moments show low error with respect to the CCSD(T)$_{r}$ data (shown in Table S9 of the SI) and, thus, provide a reliable supramolecular reference except for CO-HF and CO-HCl, where oo-pCCD-LCCD performs better, similarly to the observer for the isolated CO molecule in section~\ref{sec:co}. 

%%%%%%%%%%%%%%%%%%%%%%%%%%%%%%%%%%%%%%%%%%%%%%%%%%%%%%%%%%%%%%%%%%
%%%%%%%%%%%%  FIGURE  7                %%%%%%%%%%%%%%%%%%%%%%%%%%%
%%%%%%%%%%%%%%%%%%%%%%%%%%%%%%%%%%%%%%%%%%%%%%%%%%%%%%%%%%%%%%%%%%
\begin{figure}[h!]
    \centering
    \includegraphics[width=\columnwidth]{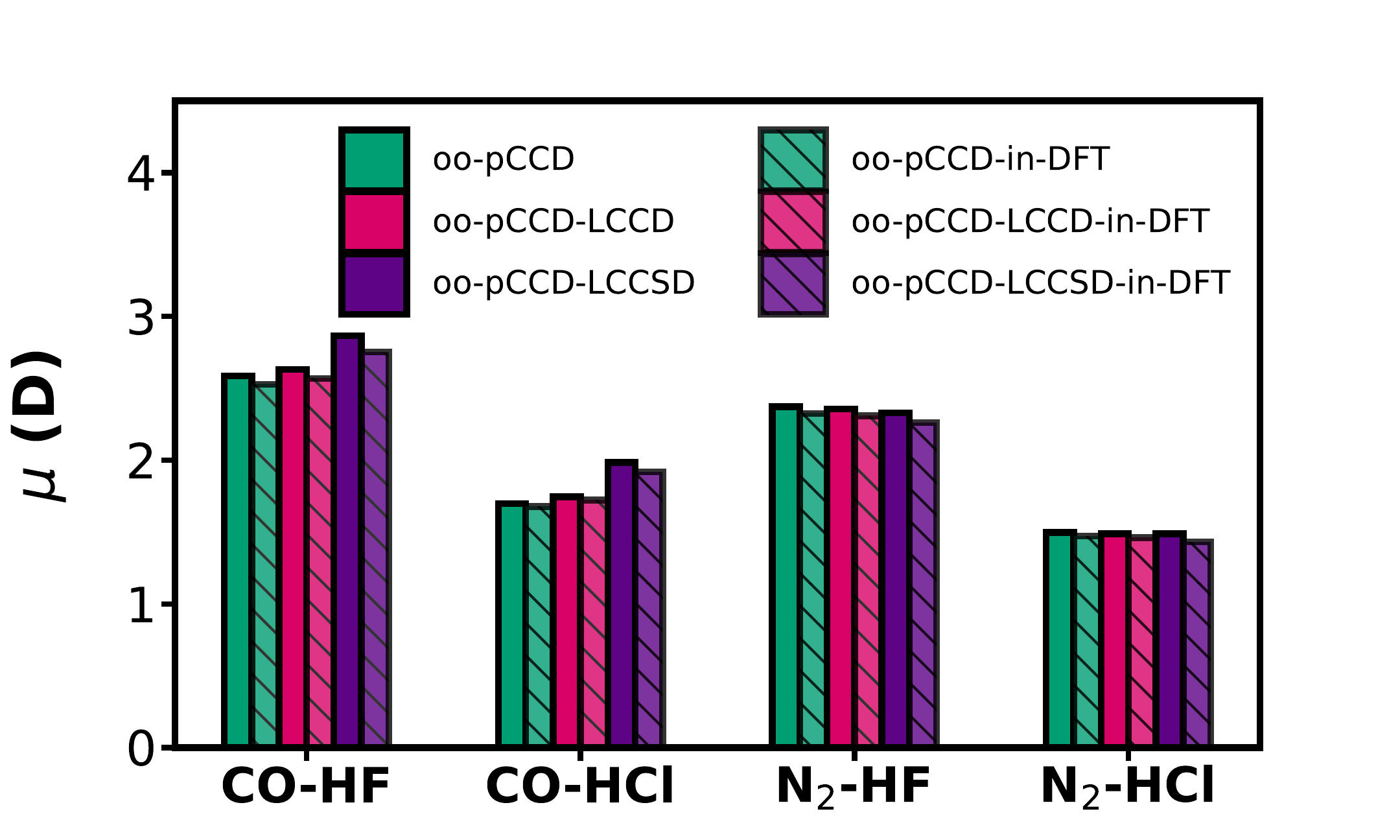}  
    \caption{Dipole moments ($\mu$ in D) of the binary complexes from oo-pCCD variants and the corresponding embedding approaches in the aug-cc-pVTZ basis set. }     
    \label{fig:complex_bar}
\end{figure}
%%%%%%%%%%%%%%%%%%%%%%%%%%%%%%%%%%%%%%%%%%%%%%%%%%%%%%%%%%%%%%%%%%
%%%%%%%%%%%%  END OF FIGURE  7         %%%%%%%%%%%%%%%%%%%%%%%%%%%
%%%%%%%%%%%%%%%%%%%%%%%%%%%%%%%%%%%%%%%%%%%%%%%%%%%%%%%%%%%%%%%%%% 

\begin{figure*}
    \centering  
    \includegraphics[width=\textwidth]{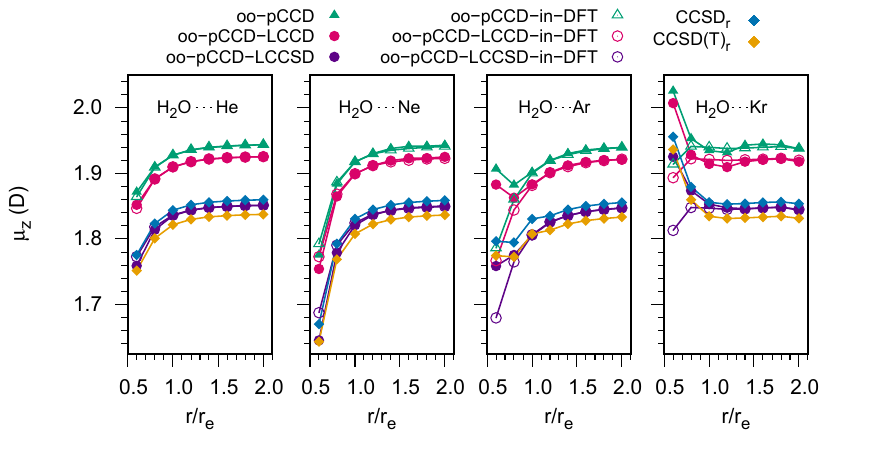}  
    \caption{Distance dependence of the calculated dipole moment components of the \ce{H2O \cdots Rg} [Rg = He, Ne, Ar, and Kr] complexes in aug-cc-pVTZ basis.}     
    \label{fig:dms_h2o_rg}
\end{figure*}

Table~\ref{tbl:complex_dipole_emb} collects dipole moments obtained from various pCCD models with and without embedding and the difference between them.
Figure~\ref{fig:complex_bar} summarizes the performance of the orbital optimized pCCD-based embedding models for dipole moments of weakly hydrogen-bonded complexes (the binary complexes, see also Figure~\ref{fig:str_complex}a).
The static embedding approach produces dipole moments closer to the respective supramolecular values with both oo-pCCD and oo-pCCD-LCC methods. 
Interestingly, the difference in embedding and supramolecular dipole moment values is lower with oo-pCCD and oo-pCCD-LCCD compared to oo-pCCD-LCCSD.
This is most likely attributed to the limitations of oo-pCCD-LCCSD when individual fragments possess multiple bonds, as we observed for the diatomics (vide supra). 
%%%%%%%%%%%%%%%%%%%%%%%%%%%%%%%%%%%%%%%%%%%%%%%%%%%%%%%%%%%%%%%%%%
%%%%%%%%%%%%  TABLE   2                %%%%%%%%%%%%%%%%%%%%%%%%%%%
%%%%%%%%%%%%%%%%%%%%%%%%%%%%%%%%%%%%%%%%%%%%%%%%%%%%%%%%%%%%%%%%%%
\begin{center}
\begin{table*}[ht!]
\setlength{\tabcolsep}{4pt}
\begin{footnotesize}
\caption {Dipole moment ($\mu$ in D) from aug-cc-pVTZ  oo-pCCD and oo-pCCD-in-DFT types of methods and their differences. The errors are calculated as $\mu_{\textrm{emb.}}-\mu_{\textrm{supra.}}$. } 
\label{tbl:complex_dipole_emb}
\begin{tabular}{lccccccccc}
\hline
{Complex}&
\multicolumn{3}{c}{oo-pCCD}&
\multicolumn{3}{c}{oo-pCCD-LCCD}&
\multicolumn{3}{c}{oo-pCCD-LCCSD}\\ 
\cline{2-4}
\cline{5-7}
\cline{8-10}
%\cline{2-4}\cline{4-7}\cline{7-10}
&\multicolumn{1}{c}{~supra.}&
\multicolumn{1}{c}{~~emb.}&
\multicolumn{1}{c}{~~error}&
\multicolumn{1}{c}{~supra.}&
\multicolumn{1}{c}{~~emb.}&
\multicolumn{1}{c}{~~error}&
\multicolumn{1}{c}{~supra.} &
\multicolumn{1}{c}{~~emb.}&
\multicolumn{1}{c}{~~error}\\
\hline
\ce{CO-HF} &~2.586 &~2.528 &-0.058 &~2.630 &~2.569&-0.061 &~2.866 &~2.752 &-0.114 \\  
\ce{CO-HCl} &~1.698 &~1.677 &-0.021 &~1.745 &~1.723 &-0.022 &~1.983 &~1.918 &-0.065\\ 
\ce{N2-HF} &~2.371 &~2.325 &-0.046&~2.357&~2.309 &-0.048 &~2.329 &~2.260 &-0.069\\    
\ce{N2-HCl} &~1.497 &~1.471 &-0.026 &~1.488 &~1.461 &-0.027 &~1.488 &~1.431 &-0.057 \\ 
\ce{H2O\cdots He} &~1.928 &~1.928 &~0.000 &~1.910 &~1.910 &~0.000 &~1.836 &~1.836 &~0.000 \\ 
\ce{H2O\cdots Ne} &~1.918 &~1.918 &~0.000 &~1.899 &~1.900 &~0.001 &~1.821 &~1.824 &~0.003  \\ 
\ce{H2O\cdots Ar} &~1.905 &~1.904 &-0.001 &~1.887 &~1.885 &-0.002 &~1.810 &~1.810 &~0.000  \\ 
\ce{H2O\cdots Kr} &~1.940 &~1.948 &~0.008 &~1.922 &~1.930 &~0.008 &~1.863 &~1.856 &-0.007  \\ 
\hline
\end{tabular}
\end{footnotesize}
\end{table*}
\end{center}
%%%%%%%%%%%%%%%%%%%%%%%%%%%%%%%%%%%%%%%%%%%%%%%%%%%%%%%%%%%%%%%%%%
%%%%%%%%%%%%  END OF TABLE   2         %%%%%%%%%%%%%%%%%%%%%%%%%%%
%%%%%%%%%%%%%%%%%%%%%%%%%%%%%%%%%%%%%%%%%%%%%%%%%%%%%%%%%%%%%%%%%%
We also study the dipole moments of the van der Waal's complexes between \ce{H2O} and the first four inert gases. 
Here, the performance of the static embedding approach is even better for all oo-pCCD variants. 
This is to be expected as, for these complexes, the electronic properties are dominated by the highly polar \ce{H2O} molecule, and it is easier to estimate them with embedding.
As far as the supramolecular results in comparison to CCSD(T)$_{r}$ are concerned, oo-pCCD-LCCSD shows the best performance, with errors comparable to CCSD$_{r}$ (bottom part of Table S9 of the SI).
Most importantly, the changes in the dipole moment with change in the inert gas molecule (decrease from He to Ar and then increase for Kr) are captured by all oo-pCCD-based methods (supramolecular and embedding).
Figure~\ref{fig:dms_h2o_rg} shows the change in dipole moments of the \ce{H2O \cdots Rg} complexes with the distance between \ce{H2O} and the inert gas atom. 
For these curves, the distance between \ce{H2O} and the Rg atom is increased in multiples of the equilibrium distances, keeping the angles the same for the respective structures.
Here, we plot the major component of the dipole, that is $\mu_z$. 
A plot for $\mu_x$ is shown in Figure S2 of the SI. 
For \ce{H2O \cdots He} and \ce{H2O \cdots Ne}, the supramolecular trends in the changes in the dipole are well-reproduced by the embedding methods throughout the distances scanned.  
For \ce{H2O \cdots Ar} and \ce{H2O \cdots Kr}, the embedding methods differ from the supramolecular variants significantly at shorter distances. 
We anticipate that this is caused by the shortcoming of the kinetic energy functional, which has been observed for other complexes with Ar and Kr.~\cite{pawel3} 
The non-parallelity errors (difference between highest error and lowest error between embedding and supramolecular curves) are 0.121, 0.116, and 0.079 (\ce{H2O \cdots Ar}), and 0.112, 0.114,  and 0.112 (\ce{H2O \cdots Kr}) for oo-pCCD-in-DFT, oo-pCCD-LCCD-in-DFT, and oo-pCCD-LCCSD-in-DFT respectively.
 
Barring the initial points for \ce{H2O \cdots Ar} and \ce{H2O \cdots Kr}, the oo-pCCD-LCCSD curves (both supra and embedding) are between those of CCSD$_r$ and CCSD(T)$_r$ for all systems. 
To conclude, the performance of both oo-pCCD-LCCSD and oo-pCCD-LCCSD-in-DFT is encouraging for these systems, keeping in mind the low computational cost of the static embedding approach. 
%%%%%%%%%%%%%%%%%%%%%%%%%%%%%%%%%%%%%%%%%%%%%%%%%%%%%%%%%%%%%%%%%%
%%%%%%%%%%%%  FIGURE 7                 %%%%%%%%%%%%%%%%%%%%%%%%%%%
%%%%%%%%%%%%%%%%%%%%%%%%%%%%%%%%%%%%%%%%%%%%%%%%%%%%%%%%%%%%%%%%%%
%%%%%%%%%%%%%%%%%%%%%%%%%%%%%%%%%%%%%%%%%%%%%%%%%%%%%%%%%%%%%%%%%%
%%%%%%%%%%%%  END OF FIGURE 7          %%%%%%%%%%%%%%%%%%%%%%%%%%%
%%%%%%%%%%%%%%%%%%%%%%%%%%%%%%%%%%%%%%%%%%%%%%%%%%%%%%%%%%%%%%%%%%

%%%%%%%%%%%%%%%%%%%%%%%%%%%%%%%%%%%%%%%%%%%%%%%%%%%%%%%%%%%%%%%%%%
%%%%%%%%                  Conclusions                   %%%%%%%%%%
%%%%%%%%%%%%%%%%%%%%%%%%%%%%%%%%%%%%%%%%%%%%%%%%%%%%%%%%%%%%%%%%%%
\section{Conclusions and outlook}\label{sec: conclusions}
%%%%%%%%%%%%%%%%%%%%%%%%%%%%%%%%%%%%%%%%%%%%%%%%%%%%%%%%%%%%%%%%%%
%%%%%%%%%%%%%%%%%%%%%%%%%%%%%%%%%%%%%%%%%%%%%%%%%%%%%%%%%%%%%%%%%%
In this work, we investigated the performance of various pCCD-based methods for predicting dipole moments. 
Our study shows that orbital optimization is essential and improves the overall performance of pCCD-based methods. 
Altogether, the best performance is obtained for the oo-pCCD-LCCD method, which is comparable to CCSD in predicting dipole moments.
Specifically, oo-pCCD-LCCD approaches CCSD accuracy in dipole moments for singly-bonded systems, while it reproduces the DMSs obtained by multi-reference methods.
Thus, we demonstrated that reliable dipole moments can also be obtained without explicitly including single excitations in the wave function model. 

For equilibrium structures, oo-pCCD-LCCD provides good agreement with the CCSD(T)$_{r}$ dipole moment values for singly-bonded systems---for instance, HF, AlF, and LiNa.
For multiply-bonded systems (such as SiO, GeS, and PN), the oo-pCCD-LCCD performance deteriorates (errors w.r.t. CCSD(T)$_{r}$ are up to around 30\%). 
The only exception is systems containing the carbon atom, where the relative errors drop below 5\%. 
The oo-pCCD-LCCD approach is also noticeably good in the modeling of DMSs.
Specifically, for the HF molecule, oo-pCCD-LCCD provides excellent agreement with FCI even in the region where CCSD (and CCSD(T)) fail. 
For carbon monoxide (up to a distance of 1.50 \AA{}), the agreement among oo-pCCD-LCCD, CCSD(T)$_{r}$, and MRCISD+Q results is remarkable.

On the contrary, the presence of linearized singles in the LCC correction on top of the pCCD reference worsens the performance when multiply-bonded diatomic molecules are considered.
That is particularly true for the investigated DMSs, where the LCCSD correction provides erroneous dipole moments. 
The presence of singles, however, improves the description of van der Waals complexes as singles are crucial for dispersion interactions.~\cite{filip-jctc-2019}
All pCCD-in-DFT models provide similar results for supramolecular and embedded dipole moments. 
As expected, for van der Waals complexes, the oo-pCCD-LCCSD provides the best agreement with coupled cluster reference data. 

Finally, this work provides a reference point for further improvements of pCCD-based models.
Specifically, our in-depth analysis of dipole moments demonstrates that when oo-pCCD provides a good reference function (like van der Waals and single-bonded systems), the LCCD (for singly-bonded systems) and LCCSD (for van der Waals interactions) corrections can improve the electric properties of the system. 
In cases where oo-pCCD is not a reliable reference model (e.g., multiple-bonded systems), LCCD does not improve the overall description, and pCCD-LCCSD tends to overcorrect dipole moments. 

It remains to be checked if using other than response density matrices (which are linear in nature) will bring some improvements.
Furthermore, it needs to be determined whether frozen-pair or tailored variants of pCCD-based models~\cite{ola-tcc} will correct for deficiencies in the investigated LCC corrections.

%%%%%%%%%%%%%%%%%%%%%%%%%%%%%%%%%%%%%%%%%%%%%%%%%%%%%%%%%%%%%%%%%%%%%%%%%%%%%%%
%%%%%%%%%%%%%%%%%%%%%%%%%%%%%%%%%%%%%%%%%%%%%%%%%%%%%%%%%%%%%%%%%%%%%%%%%%%%%%%

%%%%%%%%%%%%%%%%%%%%%%%%%%%%%%%%%%%%%%%%%%%%%%%%%%%%%%%%%%%%%%%%%%%%%%%%%%%%%%%
\section{Acknowledgments}\label{sec:acknowledgement}
%%%%%%%%%%%%%%%%%%%%%%%%%%%%%%%%%%%%%%%%%%%%%%%%%%%%%%%%%%%%%%%%%%%%%%%%%%%%%%%
%%%%%%%%%%%%%%%%%%%%%%%%%%%%%%%%%%%%%%%%%%%%%%%%%%%%%%%%%%%%%%%%%%%%%%%%%%%%%%%
%%%%%%%%%%%%%%%%%%%%%%%%%%%%%%%%%%%%%%%%%%%%%%%%%%%%%%%%%%%%%%%%%%%%%
%%%%%%%%%%%%%%%%%%%%%%%%%%%%%%%%%%%%%%%%%%%%%%%%%%%%%%%%%%%%%%%%%%%%%

%%%%%%%%%%%%%%%%%%%%%%%%%%%%%%%%%%%%%%%%%%%%%%%%%%%%%%%%%%%%%%%%%%%%%
%%%%%%%%%%%%%%%%%%%%%%%%%%%%%%%%%%%%%%%%%%%%%%%%%%%%%%%%%%%%%%%%%%%%%
R.C. and P.T.~acknowledge financial support from the OPUS research grant from the National Science Centre, Poland (Grant No. 2019/33/B/ST4/02114). P.T.~acknowledges the scholarship for outstanding young scientists from Poland's Ministry of Science and Higher Education. R.C. thanks Dr. Marta Gałynska for many interesting discussions on the topic.
\includegraphics[height=0.02\textheight]{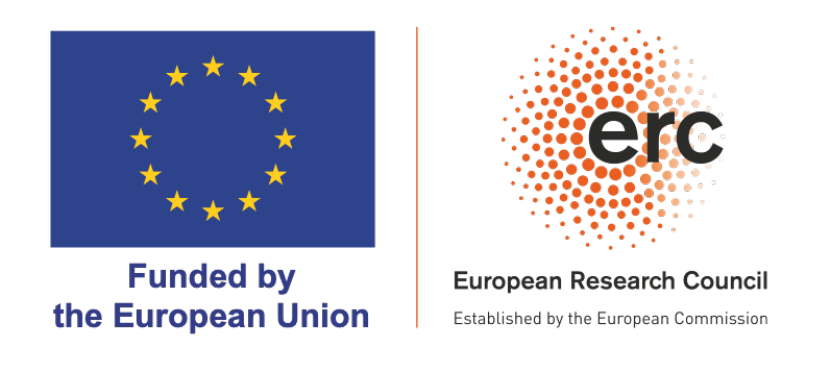} Funded/Co-funded by the European Union (ERC, DRESSED-pCCD, 101077420).
Views and opinions expressed are, however, those of the author(s) only and do not necessarily reflect those of the European Union or the European Research Council. Neither the European Union nor the granting authority can be held responsible for them. 
%%%%%%%%%%%%%%%%%%%%%%%%%%%%%%%%%%%%%%%%%%%%%%%%%%%%%%%%%%%%%%%%%%%%%
%%%%%%%%%%%%%%%%%%%%%%%%%%%%%%%%%%%%%%%%%%%%%%%%%%%%%%%%%%%%%%%%%%%%%
%%%%%%%%%%%%%%%%%%%%%%%%%%%%%%%%%%%%%%%%%%%%%%%%%%%%%%%%%%%%%%%%%%%%%
%%%%%%%%%%%%%%%%%%%%%%%%%%%%%%%%%%%%%%%%%%%%%%%%%%%%%%%%%%%%%%%%%%%%%

\section{Supplementary Information}
Experimental dipole moments and bond lengths, dipole moments from different basis sets, dipole moments without a frozen core approximation, data for DMSs of HF and CO, CO potential energy surface, structural parameters for the \ce{H2O-Rg} systems, and additional comparisons and graphs for embedding results. 

\nocite{}
\bibliography{main}% Produces the bibliography via BibTeX.

\end{document}

\renewcommand{\thefigure}{S\arabic{figure}}
\renewcommand{\thesection}{S\arabic{section}}
\renewcommand{\thetable}{S\arabic{table}}
\renewcommand{\tablename}{{Table } \hspace{-0.28cm}}

\begin{center}
\begin{spacing}{2.0}
{\LARGE\bf Toward Reliable Dipole Moments without Single Excitations: The Role of Orbital Rotations and Dynamical Correlation}
\end{spacing}

\vspace{2cm}
{\large 
{Rahul Chakraborty$^{1}$, Matheus Morato F. de Moraes$^{1}$, Katharina Boguslawski$^{1}$, Artur Nowak$^{1}$, Julian Świerczyński$^{2}$, and Pawe\l{} Tecmer$^{1}$}
}\\[4ex]

{\textit{$^{1}$Institute of Physics, Faculty of Physics, Astronomy and Informatics,\\ 
Nicolaus Copernicus University in Torun, \\
Grudzi{a}dzka 5, 87-100 Toru{n}, Poland} }\\

{\textit{$^{2}$Institute of Engineering and Technology, Faculty of Physics, Astronomy, and Informatics, Nicolaus Copernicus University in Toruń, 87-100 Toruń, Poland} }\\

\vspace{1cm}
{Corresponding authors: matheusmorat@gmail.com, k.boguslawski@fizyka.umk.pl, ptecmer@fizyka.umk.pl}

\vspace{4cm}

{\bf \Large Supplementary Information}

\vfil

\end{center}
\
\newcolumntype{M}[1]{>{\centering\arraybackslash}m{#1}}
\newpage

\newcommand\VRule[1][\arrayrulewidth]{\vrule width #1}

\begin{center}
\begin{table}[h]
\caption{Experimental bond lengths and dipole moments of diatomics studied in this work.\\} 

 \label{tab:exp_data}
    \centering
    \begin{tabular}{c|c|c|c}     
    \hline
 Type& Molecule & Bond length (\AA{}) & $\mu$ (D) \\ \hline
    \multirow{6}{*}{Singly-bonded}  &HF &~0.917 &~1.827 \\ 
    &HCl &~1.275 &~1.109  \\
    &LiH &~1.596 &~5.882 \\
    &NaH &~1.889 &~6.400  \\
    &LiF &~1.564 &~6.327  \\
    &NaCl &~2.361 &~9.002  \\ 
    &AlF &~1.654 &~1.515  \\ 
    & ClF &~1.628 &~0.850 \\
    &GaF &~1.774 &~2.400  \\  
    &LiNa &~2.810 &~0.470 \\ \hline
    
    \multirow{7}{*}{Multiply-bonded}  &CO &~1.128 &~0.112  \\
    &CS &~1.535 &~1.958  \\
    &CSe &~1.676 &~1.990 \\
    &SiO &~1.510 &~3.098 \\
    &SiS &~1.730 &~1.740 \\
    &SiSe &~2.058 &~1.100 \\ 
    &GeO &~1.625 &~3.282 \\
    &GeS &~2.012 &~2.000 \\ 
    &PN &~1.491 &~2.751  \\ \hline
    A$^{2+}$B$^{2-}$& MgO & 1.749 & 6.200 \\ \hline
    
    \end{tabular}       
\end{table}
\end{center}

\newcolumntype{P}[1]{>{\centering\arraybackslash}p{#1}}

%%%%%%%%%%%%%%%%%%% TABLE %%%%%%%%%%%%%%%%%%%%
\begin{center}
\begin{footnotesize}
\setlength{\tabcolsep}{2.3pt}
\begin{longtable}{|M{0.9cm}ccccccccccc}    
\caption {Dipole moments ($\mu$ in D) of main-group diatomics in different methods. Frozen core approximation has been used here for all the methods. CC$_u$ and CC$_r$ denote unrelaxed and relaxed dipole moments of the corresponding CC methods.}\\ 
\hline
Molecule & Basis  &pCCD &\multicolumn{1}{|p{1.5cm}|}{\centering ~pCCD\\-LCCD} &\multicolumn{1}{|p{1.5cm}|}{\centering pCCD \\-LCCSD} &CCSD$_{u}$ &CCSD(T)$_{u}$ &oo-pCCD&\multicolumn{1}{|p{1.6cm}|}{\centering oo-pCCD\\-LCCD} &\multicolumn{1}{|p{1.6cm}|}{\centering oo-pCCD\\-LCCSD} & CCSD$_{r}$ & CCSD(T)$_{r}$ \\ \hline
\endfirsthead 
\multicolumn{4}{c}%
{\tablename\ \thetable\ -- \textit{Continued from previous page}} \\ \\
\hline
Molecule & Basis  &pCCD &\multicolumn{1}{|p{1.5cm}|}{\centering ~pCCD\\-LCCD} &\multicolumn{1}{|p{1.5cm}|}{\centering pCCD \\-LCCSD} &CCSD$_{u}$ &CCSD(T)$_{u}$ &oo-pCCD&\multicolumn{1}{|p{1.6cm}|}{\centering oo-pCCD\\-LCCD} &\multicolumn{1}{|p{1.6cm}|}{\centering oo-pCCD\\-LCCSD} & CCSD$_{r}$ & CCSD(T)$_{r}$ \\ \hline
\endhead
\hline \multicolumn{5}{r}{\textit{Continued on next page}}  
\endfoot 
\hline
\endlastfoot
\multirow{6}{*}{HF}&cc-pVDZ&~1.915 &~1.899 &~1.814&~1.822 &~1.817 &~1.861 &~1.857 &~1.861 &~1.830 &~1.819 \\
&cc-pVTZ &~1.915 &~1.889 &~1.790 &~1.813 &~1.800 &~1.872 &~1.861 &~1.838 &~1.823 &~1.807\\
&cc-pVQZ &~1.907 &~1.880 &~1.789& ~1.815 &~1.801 &~1.866 &~1.852 &~1.832 &~1.826 &~1.809 \\
&aug-cc-pVDZ&~1.923 &~1.890 &~1.746 &~1.787 &~1.781 &~1.857 &~1.845 &~1.800 &~1.803 &~1.788 \\
&aug-cc-pVTZ&~1.915  &~1.880 &~1.749 &~1.794 &~1.782 &~1.860 &~1.845 &~1.806  &~1.808 &~1.791  \\
&aug-cc-pVQZ &~1.913 &~1.878 &~1.758 &~1.802 &~1.788 &~1.857 &~1.842 &~1.815 &~1.814 &~1.797 \\ \hline\hline

\multirow{6}{*}{HCl}&cc-pVDZ&~1.386 &~1.378 &~1.339 &~1.329 &~1.312 &~1.344 &~1.346 &~1.404 &~1.323 &~1.312 \\
&cc-pVTZ&~1.242 &~1.228 &~1.173 &~1.184 &~1.172 &~1.215 &~1.211 &~1.239 &~1.182 &~1.174 \\
&cc-pVQZ&~1.210 &~1.195 &~1.145 &~1.150 &~1.139 &~1.194 &~1.189 &~1.196 &~1.150 &~1.142  \\
&aug-cc-pVDZ&~1.212 &~1.199 &~1.126 &~1.144 &~1.128 &~1.198 &~1.196 &~1.190 &~1.146 &~1.132 \\
&aug-cc-pVTZ&~1.175  &~1.147 &~1.049  &~1.093 &~1.081 &~1.158 &~1.149 &~1.126 &~1.097 &~1.086  \\
&aug-cc-pVQZ&~1.163 &~1.138 &~1.061 &~1.108 &~1.096 &~1.153 &~1.146 &~1.139 &~1.110 &~1.099 \\ \hline\hline

\multirow{3}{*}{LiH}&cc-pVDZ&~5.903 &~5.857 &~5.681 &~5.735 &~5.735 &~5.721 &~5.753 &~5.951 &~5.735 &~5.735 \\
&cc-pVTZ&~5.931 &~5.889 &~5.764 &~5.848 &~5.848 &~5.829 &~5.855 &~6.015 &~5.848 &~5.848 \\
&cc-pVQZ&~5.964 &~5.924 &~5.799 &~5.855 &~5.855 &~5.854 &~5.876 &~6.025 &~5.856 &~5.856\\
\multirow{3}{*}{LiH}&aug-cc-pVDZ&~5.995 &~5.950 &~5.782 & ~5.911 &~5.911 &~5.912 &~5.936 &~6.088 &~5.912 &~5.912 \\
&aug-cc-pVTZ&~5.972 &~5.932 &~5.811 &~5.872 &~5.872 &~5.873 &~5.894 &~6.037 &~5.873 &~5.873 \\
&aug-cc-pVQZ&~5.982 &~5.944 &~5.818 &~5.868 &~5.868 &~5.869 &~5.889 &~6.031 &~5.869 &~5.869\\ \hline \hline

\multirow{6}{*}{NaH}&cc-pVDZ&~6.959 &~6.871 &~6.557 &~6.423 &~6.423 &~6.610 &~6.657 &~7.011 &~6.424 &~6.424 \\
&cc-pVTZ&~6.959 &~6.871 &~6.557 &~6.615 &~6.615 &~6.610 &~6.657 &~7.011 &~6.616 &~6.616 \\
&cc-pVQZ&~7.002 &~6.913 &~6.571 &~6.665 &~6.665 &~6.679 &~6.721 &~7.049 &~6.666 &~6.666 \\
&aug-cc-pVDZ&~7.029 &~6.940 &~6.533 &~6.678 &~6.678 &~6.679 &~6.724 &~7.065 &~6.679 &~6.679 \\
&aug-cc-pVTZ&~7.028 &~6.940 &~6.617 &~6.692 &~6.692 &~6.693 &~6.734 &~7.058 &~6.693 &~6.693 \\
&aug-cc-pVQZ&~7.027 &~6.940 &~~6.588~ &~6.687 &~6.687 &~6.696 &~6.736 &~7.052 &~6.689 &~6.689 \\ \hline \hline 

\multirow{6}{*}{LiF}&cc-pVDZ&~6.464 &~6.427 &~5.991 &~6.195 &~6.173 &~6.363 &~6.350 &~6.140 &~6.232 &~6.179 \\
&cc-pVTZ&~6.406 &~6.372 &~6.063 &~6.273 &~6.245 &~6.358 &~6.345 &~6.194 &~6.296 &~6.251 \\
&cc-pVQZ&~6.413 &~6.384 &~6.139 &~6.285 &~6.260 &~6.379 &~6.367 &~6.247 &~6.304 &~6.266 \\
&aug-cc-pVDZ&~6.490 &~6.469 &~6.250 &~6.369 &~6.359 &~6.461 &~6.452 &~6.337 &~6.389 &~6.364 \\
&aug-cc-pVTZ&~6.451 &~6.424 &~6.202 &~6.322 &~6.304 &~6.411 &~6.399 &~6.295 &~6.341 &~6.311 \\
&aug-cc-pVQZ&~6.442 &~6.415 &~6.204 &~6.324 &~6.304 &~6.407 &~6.396 &~6.317 &~6.342 &~6.310 \\ \hline \hline

\multirow{6}{*}{NaCl}&cc-pVDZ&~9.324 &~9.264 &~8.841 &~9.012 &~8.949 &~9.252 &~9.230 &~9.038 &~9.030 &~8.955 \\
&cc-pVTZ&~9.182 &~9.128 &~8.834 &~9.047 &~8.997 &~9.201 &~9.184 &~9.057 &~9.057 &~9.000 \\
&cc-pVQZ&~9.225 &~9.177 &~8.956 &~9.138 &~9.099 &~9.201 &~9.194 &~9.209 &~9.146 &~9.100 \\
&aug-cc-pVDZ&~9.350 &~9.321 &~9.088 &~9.231 &~9.199 &~9.346 &~9.339 &~9.225 &~9.242 &~9.203 \\
&aug-cc-pVTZ&~9.305 &~9.263 &~9.028 &~9.191 &~9.158 &~9.305 &~9.291 &~9.199 &~9.200  &~9.161 \\
&aug-cc-pVQZ&~9.264 &~9.224 &~9.038 &~9.179 &~9.146 &~9.278 &~9.266 &~9.203 &~9.186 &~9.148 \\ \hline \hline 

\multirow{6}{*}{AlF}&cc-pVDZ&~1.559 &~1.498 &~1.123 &~1.280 &~1.261 &~1.519 &~1.469 &~1.111 &~1.329 &~1.278 \\
&cc-pVTZ&~1.372 &~1.348 &~1.226 &~1.286 &~1.265 &~1.375 &~1.347 &~1.179 &~1.314 &~1.279 \\
&cc-pVQZ &~1.383 &~1.359 &~1.300 &~1.366 &~1.351 &~1.403 &~1.379 &~1.275 &~1.387 &~1.365 \\
&aug-cc-pVDZ&~1.433 &~1.418 &~1.352 &~1.430 &~1.431 &~1.496 &~1.475 &~1.314 &~1.458 &~1.442 \\ 
&aug-cc-pVTZ&~1.394 &~1.383 &~1.359 &~1.394 &~1.387 &~1.424 &~1.401 &~1.300 &~1.416 &~1.401 \\ 
&aug-cc-pVQZ&~1.370 &~1.342 &~1.315 &~1.400 &~1.392 &~1.409 &~1.387 &~1.325 &~1.419 &~1.406 \\ \hline \hline

\multirow{6}{*}{ClF}&cc-pVDZ&~1.303 &~1.265 &~0.962 &~1.094 &~1.073 &~1.109 &~1.091 &~1.059 &~1.128 &~1.091\\
&cc-pVTZ&~1.119 &~1.069 &~0.799 &~0.968 &~0.921 &~0.979 &~0.954 &~0.907 &~1.002 &~0.951 \\
&cc-pVQZ&~1.059 &~1.002 &~0.741 &~0.915 &~0.861 &~0.929 &~0.900 &~0.841 &~0.948 &~0.895 \\
&aug-cc-pVDZ&~1.073 &~1.013 &~0.738 &~0.884 &~0.865 &~0.875 &~0.853 &~0.832 &~0.923 &~0.890 \\ 
&aug-cc-pVTZ&~1.072 &~1.004 &~0.738 &~0.898 &~0.854 &~0.943 &~0.917 &~0.833 &~0.934 &~0.886 \\ 
&aug-cc-pVQZ&~1.072 &~1.000 &~0.730 &~0.896 &~0.845 &~0.900 &~0.872 &~0.823 &~0.929 &~0.879\\ \hline \hline

\multirow{6}{*}{GaF}&cc-pVDZ&~2.255 &~2.192 &~1.830 &~2.021 &~1.963 &~2.260 &~2.207 &~1.844 &~2.030 &~1.943 \\
&cc-pVTZ&~2.172 &~2.130 &~1.926 &~2.098 &~2.043 &~2.216 &~2.181 &~1.963 &~2.079 &~2.009 \\
&cc-pVQZ&~2.198 &~2.167 &~2.057 &~2.195 &~2.147 &~2.26 &~2.228 &~2.084 &~2.170 &~2.112 \\
&aug-cc-pVDZ&~2.301 &~2.281 &~2.187 &~2.318 &~2.290 &~2.397 &~2.370 &~2.172 &~2.307 &~2.261 \\
&aug-cc-pVTZ&~2.244 &~2.222 &~2.122 &~2.244 &~2.207 &~2.309 &~2.278 &~2.114 &~2.219 &~2.171 \\
&aug-cc-pVQZ&~2.224 &~2.192 &~2.113 &~2.251 &~2.212 &~2.295 &~2.265 &~2.138 &~2.223 &~2.176 \\ \hline \hline

\multirow{6}{*}{LiNa}&cc-pVDZ&~0.515 &~0.559 &~0.939  &~0.898 &~0.880 &~0.898 &~0.869 &~0.648 &~0.900 &~0.879 \\
&cc-pVTZ&~0.562 &~0.600 &~0.976 &~0.892 &~0.721 &~0.889 &~0.858 &~0.651 &~0.894  &~0.725 \\
&cc-pVQZ&~0.570 &~0.609 &~1.001 &~0.898 &~0.671 &~0.895 &~0.862 &~0.661 &~0.900 &~0.677 \\
&aug-cc-pVDZ&~0.558 &~0.576 &~0.846 &~0.892 &~0.864 &~0.886 &~0.857 &~0.644 &~0.895 &~0.863 \\
&aug-cc-pVTZ&~0.603 &~0.670 &~1.071 &~0.894 &~0.716 &~0.893 &~0.860 &~0.652 &~0.896 &~0.720 \\
&aug-cc-pVQZ&~0.590  &~0.646 &~1.069 &~0.898 &~0.653 &~0.901 &~0.868 &~0.661 &~0.900 &~0.661 \\ \hline \hline

\multirow{4}{*}{CO}&cc-pVDZ&~0.144 &~0.069 &~0.494 &~0.252 &~0.262 &~0.124 &~0.186 &~0.474 &~0.172 &~0.221 \\
&cc-pVTZ&~0.164 &~0.086 &~0.446 &~0.178 &~0.210  &~0.089 &~0.146 &~0.397 &~0.106 &~0.167 \\
&cc-pVQZ&~0.189 &~0.110 &~0.406 &~0.135 &~0.169 &~0.078 &~0.132 &~0.352 &~0.065 &~0.125 \\
&aug-cc-pVDZ&~0.265 &~0.165 &~0.421 &~0.172 &~0.182 &~0.073 &~0.132 &~0.386 &~0.097 &~0.141 \\
\multirow{2}{*}{CO}&aug-cc-pVTZ&~0.278 &~0.184  &~0.388 &~0.141 &~0.170 &~0.074 &~0.129 &~0.355   &~0.070 &~0.126 \\
&aug-cc-pVQZ&~0.280 &~0.191 &~0.368 &~0.128 &~0.161 &~0.078 &~0.131 &~0.344 &~0.059 &~0.117 \\ \hline\hline

\multirow{6}{*}{CS}&cc-pVDZ&~1.583 &~1.641 &~2.350 &~1.935 &~1.928 &~1.785 &~1.860 &~2.241 &~1.843 &~1.879 \\
&cc-pVTZ&~1.639 &~1.692 &~2.397 &~1.959 &~1.949 &~1.917 &~1.968 &~2.273 &~1.878 &~1.918 \\
&cc-pVQZ&~1.685 &~1.728 &~2.419 &~1.979 &~1.963 &~1.963 &~2.009 &~2.288 &~1.898 &~1.937 \\
&aug-cc-pVDZ&~1.615 &~1.677 &~2.423 &~2.021 &~2.014 &~1.851 &~1.924 &~2.323 &~1.936 &~1.971 \\ 
&aug-cc-pVTZ&~1.611 &~1.678 &~2.433 &~2.010 &~1.999 &~1.977 &~2.026 &~2.313 &~1.928 &~1.969 \\ 
&aug-cc-pVQZ&~1.626 &~1.680 &~2.416 &~2.000 &~1.984 &~1.983 &~2.028 &~2.305 &~1.919 &~1.959 \\ \hline \hline  

\multirow{6}{*}{CSe}&cc-pVDZ&~1.838 &~1.912 &~2.849 &~2.255 &~2.215 &~2.078 &~2.176 &~2.671 &~2.126 & ~2.148 \\
&cc-pVTZ&~1.825 &~1.908 &~2.911 &~2.275 &~2.233 &~2.197 &~2.274 &~2.726 &~2.153 &~2.185 \\
&cc-pVQZ&~1.857 &~1.927 &~2.927 &~2.268 &~2.222 &~2.226 &~2.297 &~2.718 &~2.146 &~2.178\\
&aug-cc-pVDZ&~1.793 &~1.890 &~2.919 &~2.335 &~2.293 &~2.167 &~2.262 &~2.741 &~2.214 &~2.234 \\
&aug-cc-pVTZ&~1.768 &~1.871 &~2.926 &~2.304 &~2.260 &~2.240 &~2.315 &~2.737&~2.183 &~2.213 \\
&aug-cc-pVQZ&~1.763 &~1.848 &~2.878 &~2.277 &~2.230 &~2.238 &~2.309 &~2.718 &~2.155 &~2.187 \\ \hline \hline

\multirow{6}{*}{SiO}&cc-pVDZ&~3.309 &~3.135 &~1.893 &~2.534 &~2.405 &~3.070 &~2.944 &~2.112 &~2.697 &~2.506 \\
&cc-pVTZ&~3.456 &~3.297 &~2.173 &~2.881 &~2.698 &~3.375 &~3.268 &~2.390 &~3.009 &~2.791 \\
&cc-pVQZ&~3.610 &~3.447 &~2.310 &~3.089 &~2.908 &~3.531 &~3.430 &~2.573 &~3.207 &~2.998 \\
&aug-cc-pVDZ&~3.613 &~3.438 &~2.162 &~2.976 &~2.850 &~3.506 &~3.398 &~2.461 &~3.115 &~2.933 \\  
&aug-cc-pVTZ&~3.662 &~3.502 &~2.301 &~3.109 &~2.943 &~3.574 &~3.472 &~2.572 &~3.232 &~3.032 \\ 
&aug-cc-pVQZ&~3.686 &~3.527 &~2.343 &~3.157 &~2.984 &~3.584 &~3.484 &~2.626 &~3.273 &~3.073 \\ \hline \hline

\multirow{6}{*}{SiS}&cc-pVDZ&~1.016 &~0.937 &~0.427 &~0.731 &~0.632 &~1.365 &~1.238 &~0.418 &~0.794 &~0.684 \\
&cc-pVTZ&~1.091 &~1.002 &~0.565 &~0.903 &~0.801 &~1.396 &~1.293 &~0.564 &~0.947 &~0.840 \\
&cc-pVQZ&~1.044 &~0.975 &~0.646 &~0.952 &~0.862 &~1.380 &~1.285 &~0.623 &~0.993 &~0.898 \\
&aug-cc-pVDZ&~1.088 &~0.995 &~0.517 &~0.853 &~0.768 &~1.433 &~1.325 &~0.532 &~0.907 &~0.815 \\
&aug-cc-pVTZ&~1.111 &~1.010 &~0.563 &~0.920 &~0.829 &~1.380 &~1.283 &~0.582 &~0.965 &~0.869 \\
&aug-cc-pVQZ&~1.106 &~1.021 &~0.638 &~0.955 &~0.869 &~1.292 &~1.212 &~0.661 &~0.997 &~0.904\\ \hline \hline 

\multirow{6}{*}{SiSe}&cc-pVDZ &~1.786 &~1.609 &~0.362 & ~0.997 &~0.840 &~2.174 &~1.983 &~0.488 &~1.167 &~0.977 \\
&cc-pVTZ&~1.786 &~1.609 &~0.362 &~1.243 &~1.064 &~2.174 &~1.983 &~0.488 &~1.388 &~1.183 \\
&cc-pVQZ&~1.797 &~1.628 &~0.465 &~1.337 &~1.166 &~2.190 &~2.007 &~0.571 &~1.473 &~1.278 \\
&aug-cc-pVDZ&~1.828 &~1.616 &~0.278 &~1.156 &~1.010 &~2.254 &~2.048 &~0.462 &~1.310 &~1.134 \\
&aug-cc-pVTZ&~1.861 &~1.652 &~0.321 &~1.261 &~1.095 &~2.146 &~1.960 &~0.501 &~1.404 &~1.212 \\
&aug-cc-pVQZ&~1.860 &~1.659 &~0.399 &~1.339 &~1.171 &~2.178 &~1.998 &~0.573 &~1.474 &~1.282 \\ \hline \hline

\multirow{6}{*}{GeO}&cc-pVDZ&~3.752 &~3.524 &~1.697 &~2.712 &~2.516 &~3.214 &~3.072 &~2.149 &~2.926 &~2.663 \\
&cc-pVTZ&~3.893 &~3.699 &~2.027 &~3.092 &~2.833 &~3.569 &~3.451 &~2.502 &~3.260 &~2.959 \\
&cc-pVQZ&~4.069 &~3.857 &~2.067 &~3.311 &~3.051 &~3.739 &~3.628 &~2.697 &~3.462 &~3.168 \\
&aug-cc-pVDZ&~4.120 &~3.921 &~2.175 &~3.287 &~3.096 &~3.805 &~3.687 &~2.669 &~3.467 &~3.210 \\
&aug-cc-pVTZ&~4.168 &~3.957 &~2.052 &~3.364 &~3.126 &~3.815 &~3.703 &~2.732 &~3.521 &~3.241 \\
&aug-cc-pVQZ&~4.174 &~3.961 &~2.043 &~3.406 &~3.157 &~3.819 &~3.709 &~2.776 &~3.553 &~3.270 \\ \hline \hline

\multirow{6}{*}{GeS}&cc-pVDZ&~2.514 &~2.325 &~0.964 &~1.730 &~1.523 &~2.900 &~2.675 &~1.141 &~1.896 &~1.672 \\
&cc-pVTZ&~2.625 &~2.432 &~1.131 &~1.993 &~1.765 &~2.956 &~2.779 &~1.298 &~2.120 &~1.887 \\
&cc-pVQZ&~2.608 &~2.436 &~1.264 &~2.116 &~1.902 &~2.978 &~2.812 &~1.419 &~2.233 &~2.014 \\
&aug-cc-pVDZ&~2.685 &~2.482 &~1.133 &~2.008 &~1.814 &~3.070 &~2.882 &~1.348 &~2.151 & ~1.944 \\
&aug-cc-pVTZ&~2.646 &~2.459 &~1.170 &~2.075 &~1.866 &~2.983 &~2.812 &~1.362 &~2.201 &~1.985 \\
&aug-cc-pVQZ&~2.647 &~2.470 &~1.250 &~2.137 &~1.929 &~2.981 &~2.819 &~1.435 &~2.253 &~2.039 \\ \hline \hline

\multirow{5}{*}{\ce{PN}}&cc-pVDZ&~2.578 &~2.486 &~1.968 &~2.394 &~2.290 &~2.493 &~2.422 &~2.155 &~2.474 &~2.368 \\
&cc-pVTZ&~2.868 &~2.736 &~2.140 &~2.638 &~2.482 &~2.704 &~2.634 &~2.377 &~2.714 &~2.582 \\ 
&cc-pVQZ&~2.984 &~2.844 &~2.247 &~2.762 &~2.595 &~2.795 &~2.726 &~2.488 &~2.833 &~2.700 \\
&aug-cc-pVDZ&~2.931 &~2.794 &~2.223 &~2.688 &~2.581 &~2.782 &~2.715 &~2.419 &~2.771 &~2.660 \\ 
&aug-cc-pVTZ&~3.023 &~2.871 &~2.240 &~2.765 &~2.614 &~2.821 &~2.752 &~2.478 &~2.843 &~2.716 \\ 
PN&aug-cc-pVQZ&~3.062 &~2.904 &~2.290 &~2.807 &~2.643 &~2.830 &~2.762 &~2.523 &~2.880 &~2.749 \\ \hline \hline

\multirow{6}{*}{MgO}&cc-pVDZ&~8.366 &~8.041 &~4.423 &~4.934 &~3.838 &~4.158 &~4.122 &~5.758 &~4.600 &~4.952 \\
&cc-pVTZ&~8.377 &~8.036 &~3.962 &~5.996 &~4.696 &~4.109 &~4.075 &~5.940 &~5.721 &~5.706 \\
&cc-pVQZ&~8.430 &~8.089 &~3.752 &~6.576 &~5.256 &~3.949 &~3.923 &~5.996 &~6.316 &~6.196 \\
&aug-cc-pVDZ&~8.366 &~8.041 &~4.423 &~6.573 &~5.341 &~4.158 &~4.122 &~5.758 &~6.349 &~6.294 \\
&aug-cc-pVTZ&~8.377 &~8.036 &~3.962 &~6.759 &~5.470 &~4.109 &~4.075 &~5.940 &~6.513 &~6.404 \\
&aug-cc-pVQZ&~8.430 &~8.089 &~3.752 &~6.911 &~5.608 &~3.949 &~3.923 &~5.996 &~6.664 &~6.510 \\ \hline \hline 

\end{longtable}
\end{footnotesize}
\end{center} 
%%%%%%%%%%%%%%%%%%% END OF TABLE %%%%%%%%%%%%%%%%%%%%

%%%%%%%%%%%%%%%%%%% TABLE %%%%%%%%%%%%%%%%%%%%
\begin{center}
\begin{table*}[ht!]
\setlength{\tabcolsep}{2.3pt}
\begin{footnotesize}
\caption{Error analysis w.r.t.~experimental values for the dataset of 20 main group diatomics studied in this work. Errors are calculated as $\lvert\mu_{\rm Method}$ - $\mu_{\rm Exp.}\rvert$}.
\begin{tabular}{|M{1cm} |M{2.05cm} |M{1cm} |M{1cm} |M{1cm} |M{1cm} |M{1.75cm} |M{1.5cm} |M{1.0cm} |M{1.0cm} |M{1cm} |M{1cm}}
\hline
\centering
Quantity & Basis  &pCCD &\multicolumn{1}{|p{1.5cm}|}{\centering ~pCCD\\-LCCD} &\multicolumn{1}{|p{1.5cm}|}{\centering pCCD \\-LCCSD} &CCSD$_{u}$ &CCSD(T)$_{u}$ &oo-pCCD&\multicolumn{1}{|p{1.6cm}|}{\centering oo-pCCD\\-LCCD} &\multicolumn{1}{|p{1.6cm}|}{\centering oo-pCCD\\-LCCSD} & CCSD$_{r}$ & CCSD(T)$_{r}$ \\ \hline
\centering
\multirow{6}{*}{\ce{MUE}}&cc-pVDZ&~0.385 &~~~~~0.325 &~~~~~0.600 &~0.303 &~0.399 &~0.334 &~~~~~0.347 &~~~~~0.493 &~0.278 &~0.313 \\
&cc-pVTZ&~0.387 &~~~~~0.316 &~~~~~0.552 &~0.181 &~0.273 &~0.339 &~~~~~0.325 &~~~~~0.409 &~0.183 &~0.204\\
&cc-pVQZ&~0.412 &~~~~~0.336 &~~~~~0.520 &~0.169 &~0.203 &~0.363 &~~~~~0.348 &~~~~~0.365 &~0.181 &~0.143 \\
&aug-cc-pVDZ&~0.426 &~~~~~0.344 &~~~~~0.498 &~0.170 &~0.222 &~0.355 &~~~~~0.337 &~~~~~0.403 &~0.164 &~0.157  \\
&aug-cc-pVTZ&~0.437 &~~~~~0.356 &~~~~~0.530 &~0.180 &~0.194 &~0.363 &~~~~~0.345 &~~~~~0.373 &~0.191 &~0.148 \\
&aug-cc-pVQZ&~0.442 &~~~~~0.363 &~~~~~0.520 &~0.193 &~0.173 &~0.376 &~~~~~0.357 &~~~~~0.349 &~0.205 &~0.143  \\ \hline

\multirow{6}{*}{\ce{RMSE}}&cc-pVDZ&~0.599 &~~~~~0.513 &~~~~~0.787 &~0.441 &~0.655 &~0.574 &~~~~~0.560 &~~~~~0.610 &~0.457 &~0.452 \\
&cc-pVTZ&~0.601 &~~~~~0.506 &~~~~~0.766 &~0.260 &~0.442 &~0.585 &~~~~~0.565 &~~~~~0.512 &~0.266 &~0.290 \\
&cc-pVQZ&~0.631 &~~~~~0.531 &~~~~~0.763 &~0.255 &~0.322 &~0.627 &~~~~~0.606 &~~~~~0.463 &~0.251 &~0.234 \\
&aug-cc-pVDZ&~0.633 &~~~~~0.530 &~~~~~0.688 &~0.269 &~0.336 &~0.607 &~~~~~0.579 &~~~~~0.505 &~0.252 &~0.258 \\
&aug-cc-pVTZ&~0.640 &~~~~~0.535 &~~~~~0.753 &~0.276 &~0.300 &~0.604 &~~~~~0.581 &~~~~~0.476 &~0.262 &~0.244 \\
&aug-cc-pVQZ&~0.651 &~~~~~0.545 &~~~~~0.764 &~0.290 &~0.271 &~0.636 &~~~~~0.614 &~~~~~0.447 &~0.276 &~0.240 \\ \hline

\end{tabular}
\end{footnotesize}
\end{table*}
\end{center} 
%%%%%%%%%%%%%%%%%%% END OF TABLE %%%%%%%%%%%%%%%%%%%%

%%%%%%%%%%%%%%%%%%% TABLE %%%%%%%%%%%%%%%%%%%%
\begin{center}
\begin{table*}[ht!]
\setlength{\tabcolsep}{2.5pt}
\caption {Dipole moment ($\mu$ in D) of diatomic molecules in aug-cc-pVQZ basis set without frozen core. CC$_u$ and CC$_r$ denote unrelaxed and relaxed dipole moments of the corresponding CC methods.}
\label{table:dipole_small}
\footnotesize
\begin{tabular}{|M{1cm} |M{1.5cm} |M{1.5cm} |M{1.5cm} |M{1.5cm} |M{1.5cm} |M{1.75cm} |M{1.5cm} |M{1.5cm} |M{1.5cm} |M{1cm} |M{1cm}}
\hline
\centering
Species  &pCCD &\multicolumn{1}{|p{1.5cm}|}{\centering ~pCCD-\\LCCD} &\multicolumn{1}{|p{1.5cm}|}{\centering pCCD- \\LCCSD} &CCSD$_{u}$ &CCSD(T)$_{u}$ &oo-pCCD&\multicolumn{1}{|p{1.75cm}|}{\centering oo-pCCD-\\LCCD} &\multicolumn{1}{|p{1.75cm}|}{\centering oo-pCCD-\\LCCSD} & CCSD$_{r}$ & CCSD(T)$_{r}$ \\ \hline

\ce{HF} &~1.913 &~1.878 &~1.764 &~1.806 &~1.792 &~1.855 &~1.840 &~1.819 &~1.818 &~1.800\\ \hline
\ce{HCl} &~1.163 &~1.138 &~1.055 &~1.104 &~1.092 &~1.151 &~1.144 &~1.133 &~1.107 &~1.095\\ \hline
\ce{LiH} &~5.984 &~5.945 &~5.798 &~5.854 &~5.851 &~5.912 &~5.927 &~5.994 &~5.856 &~5.851\\ \hline
\ce{NaH} &~7.028 &~6.946 &~6.452 &~6.577 &~6.551 &~6.740 &~6.778 &~6.960 &~6.578 &~6.553\\ \hline
\ce{LiF} &~6.444 &~6.416 &~6.186 &~6.311 &~6.288 &~6.413 &~6.399 &~6.290 &~6.328 &~6.295\\ \hline
\ce{NaCl} &~9.265 &~9.232 &~8.990 &~9.144 &~9.102 &~9.285 &~9.274 &~9.153 &~9.154 &~9.108\\ \hline
\ce{AlF}  &~1.371 &~1.345 &~1.379 &~1.439 &~1.432 &~1.394 &~1.373 &~1.374 &~1.456 &~1.446\\ \hline
\ce{ClF} &~1.072 &~1.001 &~0.744 &~0.908 &~0.853 &~0.897 &~0.868 &~0.839 &~0.940 &~0.887\\ \hline
\ce{GaF} &~2.224 &~2.188 &~2.090 &~2.270 &~2.252 &~2.295 &~2.261 &~2.117 &~2.309 &~2.267\\ \hline
\ce{LiNa} &~0.587 &~0.649 &~0.800 &~0.678 &~0.653 &~0.897 &~0.882 &~0.473 &~0.676 &~0.662\\ \hline 
\ce{CO} &~0.280 &~0.191 &~0.376 &~0.131 &~0.166 &~0.084 &~0.139 &~0.352 &~0.062 &~0.122\\ \hline
\ce{CS} &~1.629 &~1.684 &~2.469 &~2.024 &~2.006 &~1.996 &~2.043 &~2.349 &~1.939 &~1.981\\ \hline
\ce{CSe} &~1.764 &~1.850 &~2.891 &~2.252 &~2.204 &~2.266 &~2.334 &~2.710 &~2.134 &~2.170\\ \hline
\ce{SiO} &~3.687 &~3.528 &~2.378 &~3.193 &~3.013 &~3.686 &~3.576 &~2.545 &~3.304 &~3.102\\ \hline
\ce{SiS} &~1.107 &~1.022 &~0.655 &~0.976 &~0.885 &~1.364 &~1.273 &~0.630 &~1.016 &~0.922\\ \hline
\ce{SiSe} &~1.899 &~1.683 &~0.422 &~1.428 &~1.252 &~2.068 &~1.892 &~0.630 &~1.553 &~1.350\\ \hline
\ce{GeO} &~4.176 &~3.957 &~1.822 &~3.377 &~3.114 &~3.862 &~3.735 &~2.401 &~3.550 &~3.252\\ \hline
\ce{GeS} &~2.649 &~2.473 &~1.056 &~2.081 &~1.850 &~3.081 &~2.901 &~1.007 &~2.214 &~1.982\\ \hline
\ce{PN} &~3.063 &~2.907 &~2.327 &~2.839 &~2.664 &~2.826 &~2.760 &~2.559 &~2.907 &~2.773\\ \hline
\ce{MgO} &~8.430 &~8.088 &~4.139 &~6.859 &~5.526 &~3.510 &~3.551 &~5.232 &~6.660 &~6.484\\ \hline \hline

\end{tabular}
\end{table*}
\end{center} 
%%%%%%%%%%%%%%%%%%% END OF TABLE %%%%%%%%%%%%%%%%%%%%

%%%%%%%%%%%%%%%%%%% TABLE %%%%%%%%%%%%%%%%%%%%
\begin{center}
\begin{table}[h!]
\caption {Dipole moment surface of the HF molecule in aug-cc-pVTZ basis. \\}

 \label{tab:hf_dms_val}
    \centering
    \begin{tabular}{cccccc}     
    \hline
r (\AA) & oo-pCCD & oo-pCCD-LCCD & oo-pCCD-LCCSD & CCSD$_r$ & CCSD(T)$_r$ \\ \hline
0.75&-1.606&-1.594&-1.541&-1.548&-1.538\\
0.80&-1.681&-1.668&-1.616&-1.625&-1.613\\
0.85&-1.758&-1.745&-1.697&-1.703&-1.688\\
0.90&-1.835&-1.820&-1.778&-1.781&-1.765\\
0.917&-1.860&-1.845&-1.806&-1.808&-1.791\\
0.95&-1.908&-1.892&-1.862&-1.860&-1.841\\
1.00&-1.974&-1.958&-1.946&-1.938&-1.917\\
1.10&-2.080&-2.065&-2.112&-2.088&-2.060\\
1.15&-2.100&-2.089&-2.200&-2.158&-2.127\\
1.175&-2.111&-2.100&-2.239&-2.191&-2.158\\
1.20&-2.118&-2.108&-2.278&-2.223&-2.188\\
1.30&-2.134&-2.127&-2.407&-2.334&-2.291\\
1.35&-2.088&-2.085&-2.475&-2.378&-2.330\\
1.40&-2.048&-2.048&-2.524&-2.412&-2.358\\
1.45&-1.992&-1.995&-2.563&-2.436&-2.374\\
1.50&-1.921&-1.928&-2.589&-2.447&-2.377\\
1.55&-1.801&-1.812&-2.581&-2.446&-2.366\\
1.575&-1.791&-1.803&-2.605&-2.440&-2.354\\
1.60&-1.742&-1.756&-2.603&-2.430&-2.338\\
1.65&-1.637&-1.655&-2.590&-2.399&-2.294\\
1.70&-1.526&-1.547&-2.565&-2.354&-2.234\\
1.75&-1.412&-1.436&-2.528&-2.293&-2.158\\
1.80&-1.296&-1.323&-2.482&-2.217&-2.067\\
1.85&-1.182&-1.211&-2.426&-2.126&-1.963\\
1.90&-1.071&-1.102&-2.364&-2.023&-1.851\\
1.95&-0.965&-0.997&-2.298&-1.909&-1.734\\
2.00&-0.866&-0.898&-2.227&-1.786&-1.619\\
2.05&-0.773&-0.806&-2.155&-1.656&-1.512\\
2.10&-0.687&-0.720&-2.082&-1.523&-1.421\\
2.15&-0.609&-0.642&-2.009&-1.388&-1.353\\
2.20&-0.538&-0.571&-1.937&-1.253&-1.317\\
2.25&-0.474&-0.507&-1.867&-1.122&-1.320\\
2.30&-0.417&-0.449&-1.799&-0.996&-1.368\\
2.35&-0.366&-0.397&-1.734&-0.877&-1.469\\
2.40&-0.321&-0.351&-1.671&-0.765&-1.625\\
2.45&-0.281&-0.311&-1.611&-0.662&-1.838\\
2.50&-0.246&-0.275&-1.554&-0.567&-2.110\\
2.75&-0.127&-0.151&-1.311&-0.219&--- \\
3.00&-0.067&-0.086&-1.131&-0.033&--- \\
3.50&-0.022&-0.040&-0.906&--- &--- \\
4.00&-0.009&-0.030&-0.803&--- &---\\ \hline
\end{tabular}
\end{table}
\end{center}

%%%%%%%%%%%%%%%%%%% TABLE %%%%%%%%%%%%%%%%%%%%
\begin{center}
\begin{table}[h!]
\caption {Dipole moment surface of the CO molecule in aug-cc-pVTZ basis. \\}
\label{tab:co_dms_val}
    \centering
    \begin{tabular}{cccccc}     
    \hline
r (\AA) & oo-pCCD & oo-pCCD-LCCD & oo-pCCD-LCCSD & CCSD$_r$ & CCSD(T)$_r$ \\ \hline
0.75&1.291&1.295&1.240&1.224&1.223\\
0.80&1.162&1.170&1.135&1.105&1.106\\
0.85&1.014&1.027&1.025&0.972&0.978\\
0.90&0.860&0.879&0.905&0.828&0.838\\
0.95&0.697&0.722&0.781&0.674&0.691\\
1.00&0.527&0.559&0.656&0.511&0.536\\
1.05&0.352&0.392&0.534&0.342&0.378\\
1.10&0.174&0.224&0.417&0.169&0.217\\
1.125&0.085&0.140&0.362&0.081&0.137\\
1.128&0.074&0.129&0.355&0.070&0.127\\
1.13&0.068&0.123&0.351&0.063&0.120\\
1.15&-0.003&0.056&0.309&-0.008&0.056\\
1.20&-0.179&-0.107&0.213&-0.186&-0.103\\
1.25&-0.349&-0.265&0.132&-0.364&-0.259\\
1.30&-0.513&-0.415&0.069&-0.541&-0.410\\
1.35&-0.664&-0.55&0.025&-0.716&-0.555\\
1.40&-0.806&-0.675&0.006&-0.887&-0.693\\
1.45&-0.935&-0.786&0.014&-1.055&-0.821\\
1.50&-1.048&-0.879&0.066&-1.218&-0.939\\ \hline

\end{tabular}
\end{table}
\end{center}

%%%%%%%%%%%%  FIGURE     %%%%%%%%%%%%%%%%%%
\begin{figure*}
    \centering  
    \includegraphics[width=\linewidth]{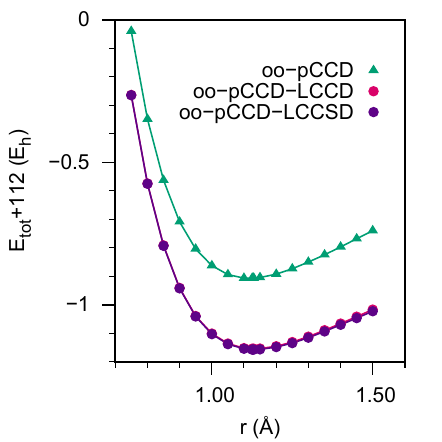}  
    \caption{Potential energy surface of CO with oo-pCCD method and its dynamic correlation correction variants in aug-cc-pVTZ basis.}     
    \label{fig:co_pes}
\end{figure*}
%%%%%%%%%%%%  END OF FIGURE   %%%%%%%%%%%%%%

%%%%%%%%%%%%%%%%%%% TABLE %%%%%%%%%%%%%%%%%%%%
\begin{center}
\begin{table}[h!]
\caption {Potential energy surface of the CO molecule in aug-cc-pVTZ basis. \\}
\label{tab:co_pes_val}
    \centering
    \begin{tabular}{cccc}     
    \hline
r (\AA) & oo-pCCD & oo-pCCD-LCCD & oo-pCCD-LCCSD \\ \hline
0.75&-112.039847&-112.263221&-112.264186\\
0.80&-112.347900&-112.573941&-112.574974\\
0.85&-112.562476&-112.791373&-112.792613\\
0.90&-112.708294&-112.940263&-112.941638\\
0.95&-112.803726&-113.038946&-113.040487\\
1.00&-112.862264&-113.100881&-113.102623\\
1.05&-112.893868&-113.136013&-113.137993\\
1.10&-112.905944&-113.151733&-113.153993\\
1.125&-112.906435&-113.154085&-113.156501\\
1.128&-112.906275&-113.154165&-113.156602\\
1.13&-112.906164&-113.15419&-113.156638\\
1.15&-112.903983&-113.153519&-113.156103\\
1.20&-112.892097&-113.145475&-113.148426\\
1.25&-112.873341&-113.130644&-113.134004\\
1.30&-112.850002&-113.111302&-113.115103\\
1.35&-112.823797&-113.089152&-113.093311\\
1.40&-112.796049&-113.065542&-113.070078\\
1.45&-112.767685&-113.041273&-113.046226\\
1.50&-112.739429&-113.016702&-113.022521\\
\hline
\end{tabular}
\end{table}
\end{center}

%%%%%%%%%%%%%%%%%%% TABLE %%%%%%%%%%%%%%%%%%%%
\begin{center}
\begin{table}[h]
\caption {Equilibrium bond parameters of the \ce{H2O\cdots Rg} [Rg = He, Ne, Ar, Kr] complexes used in this work. \\} 

 \label{tab:h2o_rg_geom}
    \centering
    \begin{tabular}{c|c|c|c}     
    \hline
    Species & r$_e$ (\AA) & $\theta_e$ \\ \hline 
    \ce{H2O\cdots He} &~3.186 & ~79\degree \\ \hline
\ce{H2O\cdots Ne} &~3.228 &~76\degree \\ \hline
\ce{H2O\cdots Ar} &~3.651 &~100\degree \\ \hline
\ce{H2O\cdots Kr} &~3.884 &~109.3\degree \\ \hline
      \end{tabular}       
\end{table}
\end{center}
%%%%%%%%%%%%%%%%%%% END OF TABLE %%%%%%%%%%%%%%%%%%%%

%%%%%%%%%%%%%%%%%%% TABLE %%%%%%%%%%%%%%%%%%%%
\begin{center}
\begin{table*}[ht!]
\setlength{\tabcolsep}{6pt}
\begin{footnotesize}
\caption {Supramolecular dipole moments ($\mu$ in D) of the binary complexes in aug-cc-pVTZ basis. CC$_u$ and CC$_r$ denote unrelaxed and relaxed dipole moments of the corresponding CC methods. The errors with respect to reference CCSD(T)$_r$ values are calculated as $\epsilon= \mu_{\textrm{Method}}-\mu_{\textrm{CCSD(T)$_{r}$}}$.}
\label{table: complex_dipole_benchmark}

\begin{tabular}{|M{1.5cm} |M{2cm} |M{1cm} |M{2cm} |M{1cm} |M{2cm} |M{1cm} |M{1cm} |M{1cm} |M{1cm} |M{1cm}}
\hline
Species&\multicolumn{1}{|p{2cm}|}{\centering oo-pCCD \\ }  &$\epsilon$ &\multicolumn{1}{|p{2cm}|}{\centering oo-pCCD- \\ LCCD }   &\centering $\epsilon$ &\multicolumn{1}{|p{2cm}|}{\centering oo-pCCD-\\ LCCSD}   &$\epsilon$ & CCSD$_{r}$ & $\epsilon$& CCSD(T)$_{r}$ \\ 
\hline

\ce{CO-HF} &~2.586 &-0.030 &~2.630 &-0.014 &~2.866 &-0.250 &~2.571 &-0.045  &~2.616 \\ \hline
\ce{CO-HCl} &~1.698  &-0.007  &~1.745  &~0.040 &~1.983 &~0.278 &~1.652 &-0.053 &~1.705 \\ \hline
\ce{N2-HF} &~2.371 &~0.047  &~2.357  &~0.043 &~2.329 &~0.005 &~2.335 &~0.011 &~2.324 \\ \hline
\ce{N2-HCl} &~1.497 &~0.064   &~1.488   &~0.055 &~1.466 &~0.033 &~1.441 &~0.008 &~1.433 \\ \hline
\ce{H2O\cdots He} &~1.928 &~0.107 &~1.910 &~0.089 &~1.835 &~0.032 &~1.844 &~0.023&~1.821 \\ \hline 
\ce{H2O\cdots Ne}  &~1.918 &~0.110 &~1.899 &~0.091 &~1.821 &~0.013 &~1.831 &~0.023&~1.808  \\ \hline 
\ce{H2O\cdots Ar}  &~1.905 &~0.104&~1.887 &~0.086 &~1.810 &~0.009 &~1.822 &~0.021&~1.801  \\ \hline 
\ce{H2O\cdots Kr}  &~1.940 &~0.096 &~1.922 &~0.078 &~1.863 &~0.019 &~1.865 &~0.021&~1.844  \\ \hline \hline \\ \\ \\
\end{tabular}
\end{footnotesize}
\end{table*}
\end{center}
%%%%%%%%%%%%%%%%%%% END OF TABLE %%%%%%%%%%%%%%%%%%%%

%%%%%%%%%%%%  FIGURE     %%%%%%%%%%%%%%%%%%
\begin{figure*}[ht!]
    \centering  
    \caption{Distance dependence of the calculated dipole moment components in aug-cc-pVTZ basis for the \ce{H2O \cdots Rg} [Rg = He, Ne, Ar, and Kr] complexes.}   
    \includegraphics[width=\linewidth]{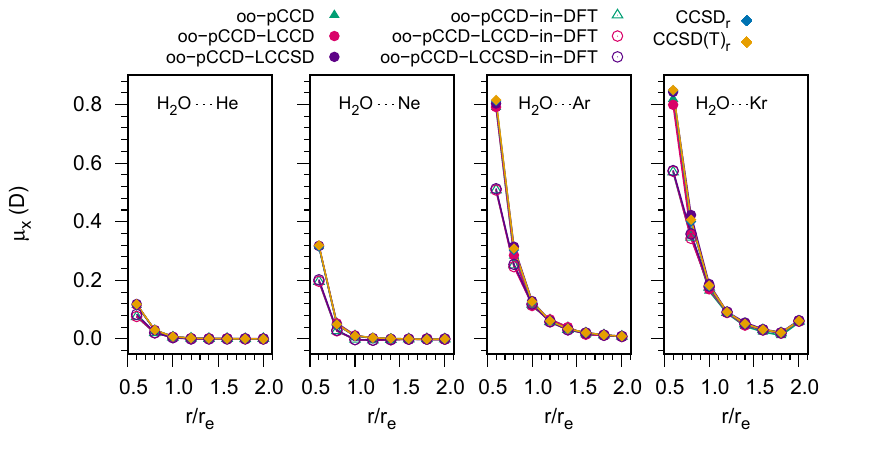}    
    \label{fig:dms_h2o_rg}
\end{figure*}
%%%%%%%%%%%%  END OF FIGURE   %%%%%%%%%%%%%%

\FloatBarrier
\newpage
\printbibliography[
heading=bibintoc,
title={References}
]